\documentclass[12pt]{iopart}
\usepackage{graphicx,caption,subcaption,float,textcomp,siunitx,mathfixs}
\usepackage{bm}
\usepackage{natbib}
\renewcommand{\vec}[1]{\boldsymbol{#1}}

\begin{document}

\bibliographystyle{abbrvnat}

\title[Acousto-elasticity of Skeletal Striated Muscle]{Acousto-elasticity of Transversely Isotropic Incompressible Soft Tissues: Characterization of Skeletal Striated Muscle}

\author{Jean-Pierre Remeni\'eras$^{1*}$, Mah\'e Bulot$^1$, Jean-Luc Gennisson$^2$, Fr\'ed\'eric Patat$^{1,3}$, Michel Destrade$^4$ and Guillaume Bacle$^{1,5}$ }
\address{$^1$UMR 1253, iBrain, Universit\'e de Tours, Inserm, Tours, France.}
\address{$^2$Laboratoire d'imagerie biom\'edicale multimodale à Paris-Saclay, Universit\'e Paris-Saclay, CEA, CNRS UMR 9011, INSERM UMR 1281, France.}
\address{$^3$Inserm CIC-IT 1415, Tours, France.}
\address{$^4$School of Mathematics, Statistics and Applied Mathematics, NUI Galway, University Road, Galway, Ireland.}
\address{$^5$Service de chirurgie orthopédique 1A, Unité de chirurgie de la main et du membre sup\'erieur, CHRU de Tours }
\ead{jean-pierre.remenieras@univ-tours.fr}

\vspace{10pt}

\begin{abstract}

Using shear wave elastography, we measure the changes in the wave speed with the stress produced by a striated muscle during isometric voluntary contraction. 
To isolate the behaviour of an individual muscle from complementary or antagonistic actions of adjacent muscles, we select the \textit{flexor digiti minimi} muscle, whose sole function is to extend the little finger. 
To link the wave speed to the stiffness, we develop an acousto-elastic theory for shear waves in homogeneous, transversely isotropic, incompressible solids subject to an uniaxial stress. 
We then provide measurements of the apparent shear elastic modulus along, and transversely to, the fibre axis for six healthy human volunteers of different age and sex.
The results display a great variety across the six subjects. 
We find that the slope of the apparent shear elastic modulus along the fibre direction changes inversely to the maximum voluntary contraction (MVC) produced by the volunteer. 
We propose an interpretation of our results by introducing the S (slow) or F (fast) nature of the fibres, which harden the muscle differently and accordingly, produce different  MVCs.
This work opens the way to measuring the elastic stiffness of muscles in patients with musculoskeletal disorders or neurodegenerative diseases. 

\end{abstract}

\vspace{2pc}

\date{today}

%\submitto{\PMB}
\maketitle

%%%%%%%%%%%%%

\section{Introduction}

%%%%%%%%%%%%%%

Affecting around $25\%$ of people worldwide, musculoskeletal disorders have a high prevalence in the adult population, coupled to  enormous and increasing health and societal impacts \citep{Adams1995,Woolf2001,ScientificGroup2003,Badley1994}. Although mainly non-lethal, these pathologies cause significant morbidity with decreased function in daily life activities and lower quality of life \citep{Vos2012,Reginster2002,Storheim2014}. They also generate significant economic costs \citep{Jacobson1996,ScientificGroup2003}. 

The pathophysiology of many of these disorders is still not completely understood and the development of new diagnostic strategies and bio-markers specific to musculoskeletal tissues is crucial to medical progress \citep{Storheim2014}. 
Also, skeletal, voluntary-controlled, muscles play a big role in motorizing joints, maintaining posture, and regulating peripheral blood flow.
Hence, innovative assessments of muscle mechanical properties and dynamics linked to its very specific structural fibrillary organization can improve our understanding of normal and pathological muscle tissue behaviour and strength \citep{Gijsbertse2017,Storheim2014}. 

Over the past twenty years, ultrasound imaging techniques have gained sufficient temporal resolution to become ultra-fast ($>$1000 frames/s)  and investigate the kinematics of muscle.
Hence, they have been used to assess the dynamical behaviour and structural changes of normal and pathological contractile tissues \citep{Deffieux2008,Downs2018,Eranki2013,Lopata2010,Loram2006,D'hooge2000,Yeung1998,Miyatake1995,Nagueh1998}.

Nonetheless, the biomechanical characteristics of skeletal muscle remain difficult to clarify fully  because of its complex structural organization and its contractile properties \citep{Gennisson2010}. Indeed, as can be seen to the naked eye, skeletal muscle tissue is composed of families of parallel muscular fibres.
Muscle contraction is carried out by the shortening of these fibres, which results from the active sliding of the thick myosin filaments between the fine actin filaments found within the fibres. 
Therefore, the main biomechanical characteristics of muscle tissue associated with contraction are shortening and hardening \citep{Ford1981}. Thus, techniques that provide quantitative data on tissue deformation and elastic properties could be of great help in understanding dynamic muscle behaviour. One such technique is quantitative Shear Wave Elastography (SWE), proposed by \citet{Sarvazyan1998}, and then refined and used to quantitatively characterize the mechanical parameters of normal skeletal muscle tissues \citep{Gennisson2010,Bouillard2011,Koo2014,Nordez2008,Nordez2010,Tran2016}. Some attempts have also been made to indirectly evaluate muscle forces based on muscle elasticity using SWE \citep{Bouillard2011,Kim2018,Hug2015}.

Interestingly,  muscle stiffness increases differently with tension during sustained contraction, depending on the type of motor units activated, according to \citet{Petit1990}, who performed measurements in the \emph{peroneus longus} muscle of anesthetized cats. 
These authors found that the stiffness/tension slope is greater when (slow) S-type motor units are activated, compared to (fast fatigue-resistant) FR-type and (fast fatiguable) FF-type motor units. Their result suggests that S-type motor units contribute more to muscle hardness during contraction than F-type ones, and that the stiffness/tension relationship must consequently change according to the S/F ratio. 

During voluntary contraction, an axial stress is induced inside the muscle tissue by the shortening of the fibres which modifies its mechanical properties. The goal of this paper is to measure experimentally the changes in shear wave speed during voluntary contraction on healthy volunteers and to model these changes with the acousto-elasticity theory. This theory couples nonlinear elasticity modelling of materials and elastic wave propagation, and links the wave speed to uni-axial stress using high-order elastic constants.
Due to the presence of fibres, muscles are considered as anisotropic, specifically transversely isotropic (TI). It follows that shear waves propagate at different speeds depending on the orientation of the propagation and polarization directions with respect to the fibre axis \citep{Gennisson2003}.
We show in the next section how acousto-elasticity theory can be adapted to study shear wave propagation in an homogeneous TI incompressible solid, subject to a uniaxial stress, extending the available theory for isotropic solids \citep{Gennisson2007,Destrade2010third}. 
Acousto-elasticity theory links the shear wave speed to the uniaxial stress \citep{Gennisson2007} or, equivalently, to the uni-axial elongation \citep{Destrade2010third}. 
Both formulations have been used for \textit{in vivo} experiments when the stress is applied directly by pressing the ultrasound probe onto the tissue \citep{Latorre2012, jiang2015measuring, Bernal2015, Otesteanu2019,Bayat2019}. This approach was also developed for TI media \citep{bied2020acoustoelasticity}.
Here we measured the stress directly with a force sensor \citep{Bouillard2014} applied on the  \emph{flexor digitimi minimi} muscle.

%%%%%%%%%%%%%%%%%%%%%%%%%%%%%%%%%%%%%%%%%

\section{Acousto-elasticity in fibre muscle}
\label{AE theory}

%%%%%%%%%%%%%%%%%%%%%%%%%%%%%%%%%%%%%%%%%

%++++++++++++++++++++++++++++++++++++++++++++++++++++++

\subsection{\label{Background} Uniaxial stress in incompressible transversely isotropic solids}

%++++++++++++++++++++++++++++++++++++++++++++++++++++

We model muscles as soft incompressible materials with one preferred direction, associated with a family of parallel fibres.

TI \emph{compressible} solids are described by five independent constants, for example the following set \citep{Rouze2020}: $\mu_{\rm L},E_{\rm L}, E_{\rm T}, \nu_{\rm TT},\nu_{\rm LT}$, 
where $ \mu_{\rm L}$ is the shear elastic modulus relative to deformations along the fibres, $E_{\rm L}$, $E_{\rm T}$ are the Young moduli along, and transverse to, the fibres, respectively, and $\nu_{\rm TT}$, $\nu_{\rm LT}$ are the Poisson ratios in these directions. 
The shear elastic modulus $\mu_{\rm T}$ relative to the transverse direction is $\mu_{\rm T}=\frac{ E_{\rm T}}{2\left(1+ \nu_{\rm TT}\right)}$. 

For \emph{incompressible} TI materials, there is no volume change. 
This constraint leads to the following relations (see \cite{Rouze2020} for details),
\begin{equation} 
\nu_{\rm LT} = \textstyle \frac{1}{2}, \qquad \nu_{\rm TT}=1-\frac{E_{\rm T}}{2E_{\rm L}}. \label{eq3} 
\end{equation}
Thus, only three independent constants are required to fully describe a given transversely isotropic, linearly elastic, incompressible solid. 
Here we choose the three material parameters $\mu_{\rm T}$, $\mu_{\rm L}$, and $E_{\rm L}$, as proposed by \cite{Li2016}. 
Note that other, equivalent choices can be made \citep{Chadwick1993,Rouze2013,Papazoglou2006}.

We call $x_{\rm 1}$ the axis along the fibres and $\sigma_{\rm 11}$ the uniaxial stress applied by the volunteers in that direction during the voluntary contractions. 
The resulting extension in that direction is $e$ ($e>0$: elongation, $e<0$: contraction).
Then a simple analysis \citep{Chadwick1993} shows that $\sigma_{\rm 11} = E_{\rm L}e$, as expected.

%+++++++++++++++++++++++++++

\subsection{\label{Background2} Third-order expansion of the strain energy in a TI incompressible solid}

%++++++++++++++++++++++++++++

Acousto-elasticity calls for a third-order  expansion of the elastic strain energy $W$ in the powers of $\bm{E}$, the Green-Lagrange strain tensor. 
For transversely isotropic incompressible solids, the expansion can be written as \citep{Destrade2010second},
\begin{equation}
 W = \mu_{\text  T}I_{\rm2} + \alpha_1 I_{\rm4}^2+\alpha_2I_{\rm5} +\frac{A}{3}I_{\rm3}+\alpha_3I_{\rm2}I_{\rm4}+\alpha_4I_{\rm4}^3+\alpha_5I_{\rm4}I_{\rm5},
\label{W} 
\end{equation}
where the second-order elastic constants $\alpha_1$, $\alpha_2$ are given by 
\begin{equation}
\alpha_1=\textstyle\frac{1}{2}\left(E_{\rm L}+\mu_{\rm T}-4\mu_{\rm L}\right),
\qquad \alpha_2=2\left(\mu_{\rm L}-\mu_{\rm T}\right),
\label{ConstOrdr2} 
\end{equation}
and $A$, $\alpha_3$, $\alpha_4$ and $\alpha_5$ are third-order elastic constants. 
The strain invariants used in (\ref{W}) are
\begin{equation}
I_2 = \tr(\bm{E}^2), \quad
I_3 = \tr(\bm{E}^3), \quad
I_4 = \bm{A \cdot EA}, \quad
I_5 = \bm{A \cdot E}^2 \bm{A},
\label{InvariantI} 
\end{equation}
where $\vec{A}$ is the unit vector in the fibres direction when the solid is unloaded and at rest.
Note that \cite{Li2020} call $\alpha_1$  the  $C_{\rm qSV}$ parameter, because it quantifies the spatial dependence of the  speed $v_{\rm qSV}$ for the quasi shear vertical mode wave in an undeformed TI solid. 
\cite{Li2020} show that $\alpha_1$ can be negative or positive (with $2\alpha_1>-4 \mu_{\rm L}$, because $E_\text{L} + \mu_\text{T}>0$).

For isotropic third-order elasticity, \cite{Gennisson2007} measured the parameter $A$ for soft phantom gels and found that it can be positive or negative even for solids which have a similar second-order shear modulus $\mu$. Hence they found  $\mu=8.5$ kPa, $A = - 21.5$ kPa for a Gelatin-Agar phantom gel, and $\mu = 8.1$ kPa, $A = +10.7$ kPa for a PVA phantom gel. Thus there is a important difference from the nonlinear point of view between these two kinds of material even if their linear shear modulus are quite similar. 
As we will see, this remark carries over to TI muscle,  where the hardening effect with effort proves to be much more important than the stiffness at rest.

%++++++++++++++++++++

\subsection{\label{actotho} Elastic waves in incompressible TI solids under uni-axial stress}

%++++++++++++++++++++

We now study the propagation of small-amplitude plane body waves in a deformed, TI incompressible soft tissue.
\cite{Destrade2010second} or \cite{Ogden2011} show that it is equivalent to solving a $2\times 2$ eigenproblem for the acoustical (symmetric) tensor. 
Its eigenvectors are orthogonal and give the two possible directions of transverse polarization; its eigenvalues are real and give the corresponding wave speeds.

\begin{figure}%[H]
    \centering
        \includegraphics[width =0.9 \textwidth]{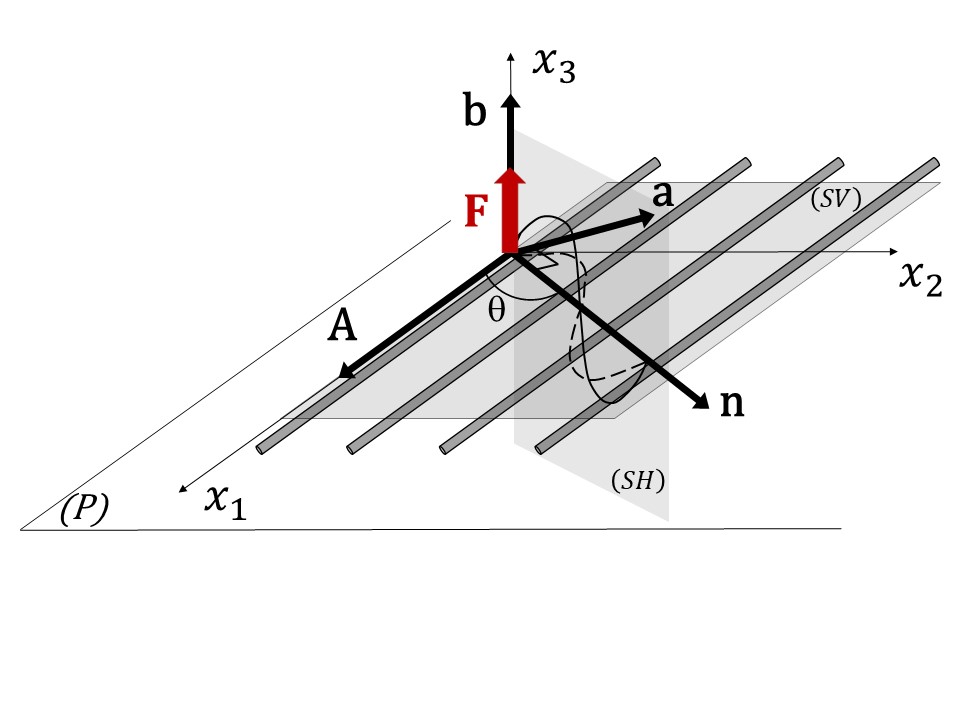}
       \caption{
       {\small
       $(P)$ is the ($\bm{A},\bm{n}$)-plane where $\bm{A}$ is a unit vector in the fibers direction when the solid is unloaded and at rest, and $\bm{n}$ is a unit vector in the direction of propagation. 
       The solid is subject to a uni-axial tensile stress $\sigma_{11}$ applied along the fibres. Two purely transverse waves propagate in an incompressible transversely isotropic solid: the shear-vertical (SV) mode with  polarization $\bm{a}$ in the $(\bm{A},\bm{n})-$plane, and the shear-horizontal (SH) mode with polarization $\bm{b}$ in the $(\bm{A}\times \bm{n},\bm{n})-$plane. Here $\theta$ is the angle between $\bm{n}$ and $\bm{A}$. In our experiments, the radiation force $\bm{F}$ is applied along the $x_{\rm 3}$ axis, and we measure the speed of waves travelling along the fibres ($\theta=0^\circ$) and transverse to the fibres ($\theta=90^\circ$). Ultrasound tracking measures the $x_{\rm 3}$ component of the shear wave displacement and is sensitive only to the (SH) propagation mode.}}
        \label{fig:Waves_SHSV}
\end{figure}
    
One eigenvector is $\bm{b}=\bm{A}\times \bm{n}$, orthogonal to both the fibres and the direction of propagation $\bm{n}$ (see Figure 1).
It corresponds to the shear-horizontal (SH) wave mode. The second one is $\bm{a}=\bm{b}\times\bm{n}$ which lies in the shear-vertical (SV) plane. 
Calculations of their wave speed $v$ as a function of the uni-axial stress $\sigma_{\rm 11}$, the propagation angle $\theta$ and the second and third order elastic moduli are detailed in the supplementary file.
In our experiments, the radiation force $\bm{F}$ used to induce the transient shear wave is applied along the $x_{\rm 3}$ axis. Ultrasound tracking measures the $x_{\rm 3}$ component of the shear wave displacement and is sensitive only to the (SH) propagation mode.

We may introduce the non-dimensional coefficients of nonlinearity $\beta_\parallel$ and $\beta_\perp$ as
\begin{eqnarray}
\beta_\parallel &= 1+\frac{1}{E_{\rm L}}\left(\mu_{\rm L}-\mu_{\rm T}+\frac{A}{4}+\alpha_3+\frac{\alpha_5}{2}\right),\label{BetaParr}\\
\beta_{\perp}&=\frac{1}{E_{\rm L}}\left(3\mu_{\rm T}+\frac{A}{2}-\alpha_3\right),
\label{BetaPerp}
\end{eqnarray}
to write the acousto-elasticity equation of the (SH) mode as follows 
\begin{equation}
\rho_{\rm 0}v^2=\left(\mu_{\rm L}-\beta_\parallel\sigma_{\rm 11}\right)\cos^2\theta
+\left(\mu_{\rm T}+\beta_\perp\sigma_{\rm 11}\right)\sin^2\theta.\label{resultSH}
\end{equation}

Note that this equation correspond to (3) and (9) proposed with a different approach by \cite{bied2020acoustoelasticity} for special cases $\theta=0$° (propagation \emph{along the fibres}) and $\theta=90$° (propagation \emph{transversely to the fibres}).
Here $\rho_{\rm 0} $ is the mass density, which remains constant throughout the deformation because of incompressibility.
In this paper, we take  $\rho_{\rm 0} =1000$ kg/m$^3$, because most human soft tissues are assumed to have the same density as water.
Notice that neither speed depends on the third-order constant $\alpha_4$, and that the speed of waves travelling transversely to the fibres does not depend on $\mu_{\rm L}$ and $\alpha_5$ either. 
However, the longitudinal Young modulus $E_{\rm L}$ does appear in that speed's expression, showing the interplay of axial and transverse linear parameters in the acousto-elastic effect. 

%%%%%%%%%%%%%%%%%%%

\section{Materials and methods}

%%%%%%%%%%%%%%%%%%%

%++++++++++++++++++++++

\subsection{Study purpose}

%++++++++++++++++++++++

Our goal is  to measure the changes in the muscle stiffness, as measured by  $\rho v^2$ for the (SH) waves, as a function of the stress $\sigma_{11}$ produced by a striated muscle during isometric contraction. 

For this purpose, we use  the Shear Wave Elastography (SWE) method, as provided by the Supersonic Shear Imaging technique included into the Aixplorer Imaging System (Supersonic Imagine, Aix en Provence, France, version V12.3). 
In principle, shear viscosity, which is frequency-dependent, is expected to modify the shear wave speed measured by the SWE technique. However, if the shear viscosity is small compared to the shear elastic modulus, the dispersion effect is limited and the muscle can be considered as a purely elastic medium.  
Moreover, as shown by \cite{bercoff2004role}, the effect of soft tissue viscosity on the shear wave speed is small provided the attenuation length is much larger than the wavelength.

We call $\mu = \rho_{\rm 0}v^2$ the ``apparent shear modulus''.
In our \textit{in vivo} study, we measure the changes in $\mu_{\parallel} = \mu_{\parallel}(\sigma_{11})$ along the fibre direction and the changes in $\mu_{\bot}=\mu_{\bot}(\sigma_{11})$ transversely to the fibre direction, with the axial stress $\sigma_{11}$ produced by the muscle. 
Then we use inverse analysis to link these experimental results to the acoustic-elasticity theory developed in Section \ref{AE theory}. 
We carry an {\it in vivo} feasibility study on six healthy volunteers with different age and sex. 

\begin{figure}%[H]
    \centering
        \includegraphics[width = 0.49\textwidth]{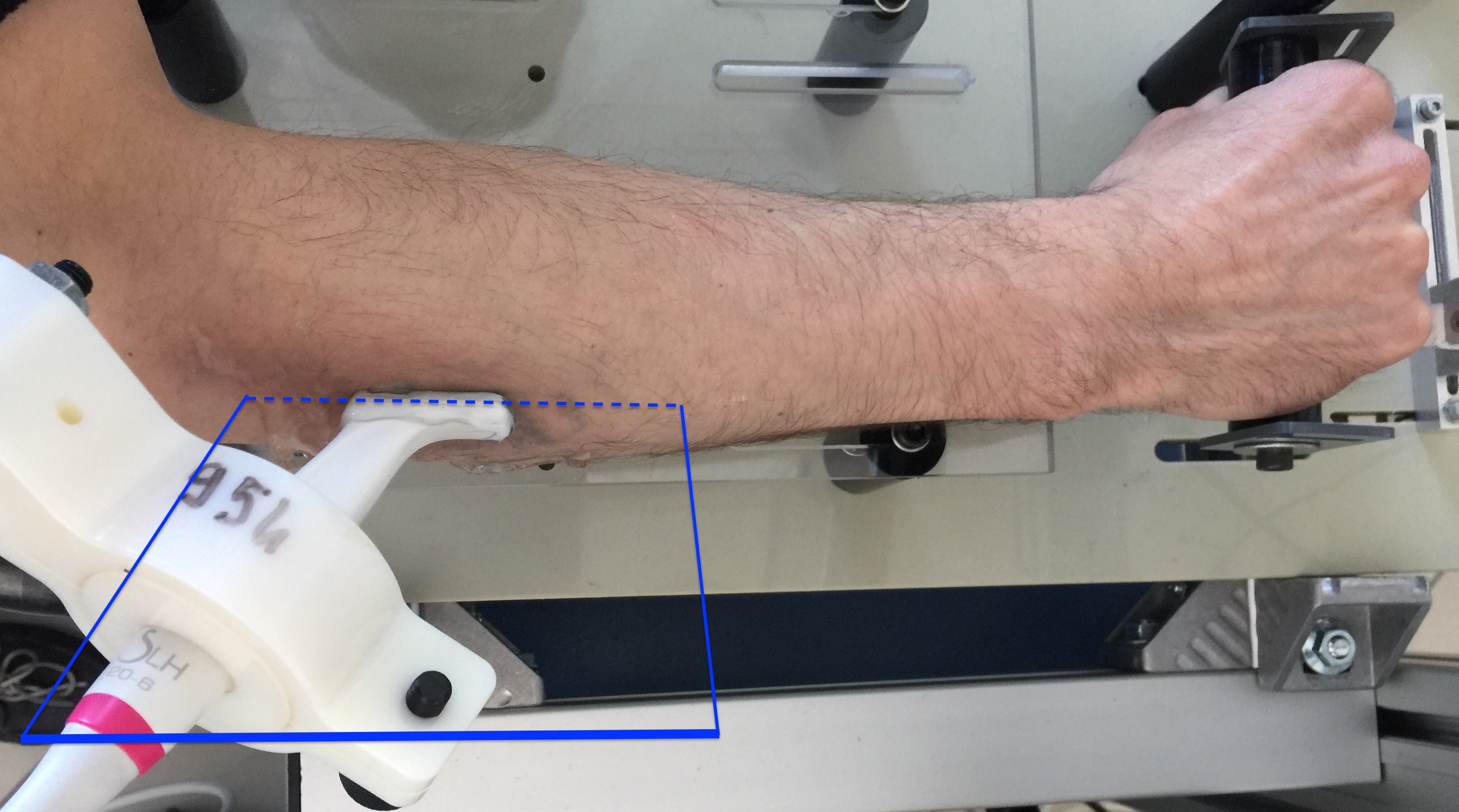}
        \includegraphics[width =0.49 \textwidth]{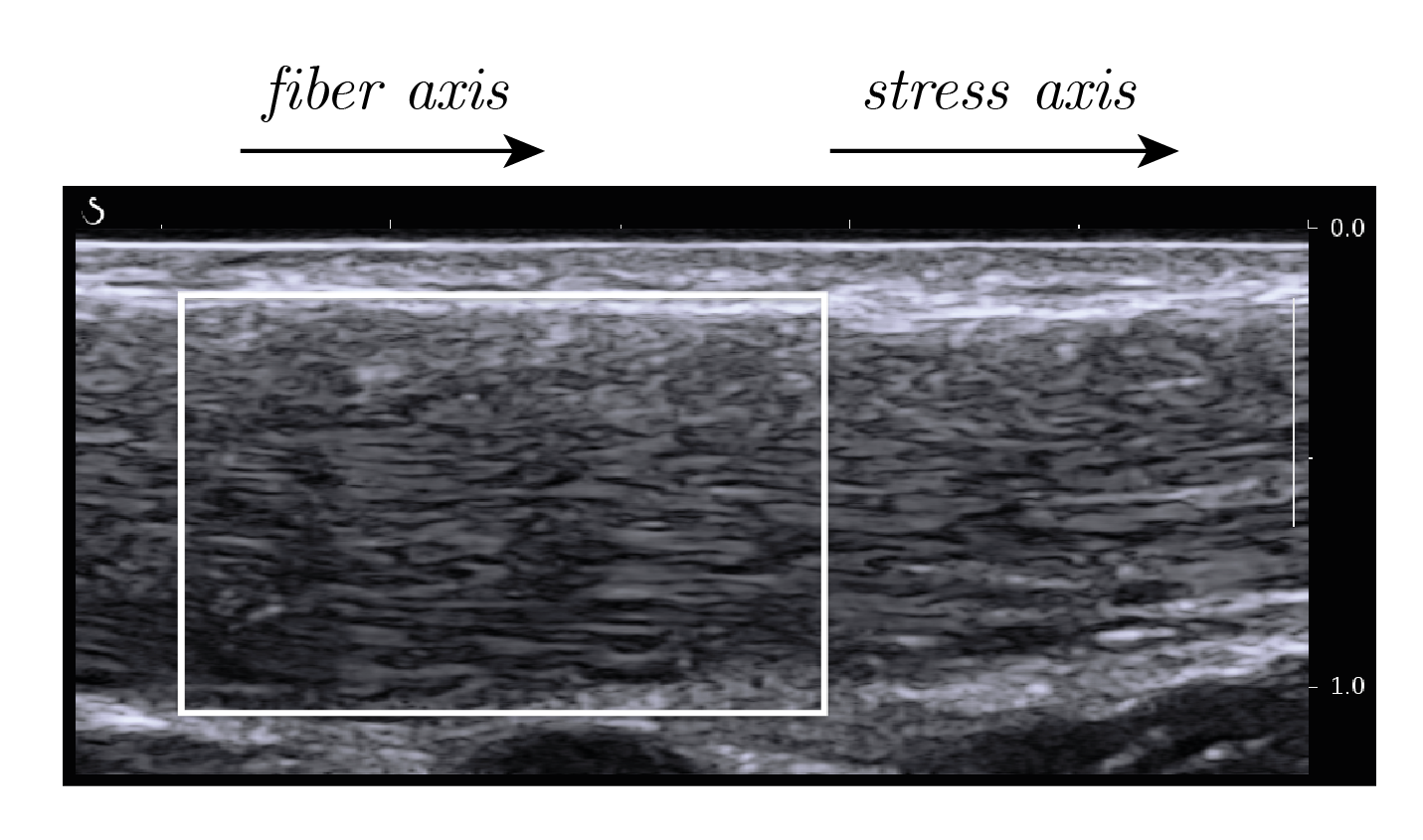}
        \caption{
          {\small
          (a) SLH20-6 musculoskeletal probe placed above the right hand \textit{flexor digiti minimi} muscle in the fibre direction. (b) Bmode image (26.7 $\times$ 14.7 mm) in the probe plane showing the fibres. The muscle studied is at a depth  between 0.5 and 1 cm.}}
           \label{BmodeLongi}
    \end{figure}
    
\begin{figure}%[H]
          \centering
        \includegraphics[width = 0.49\textwidth]{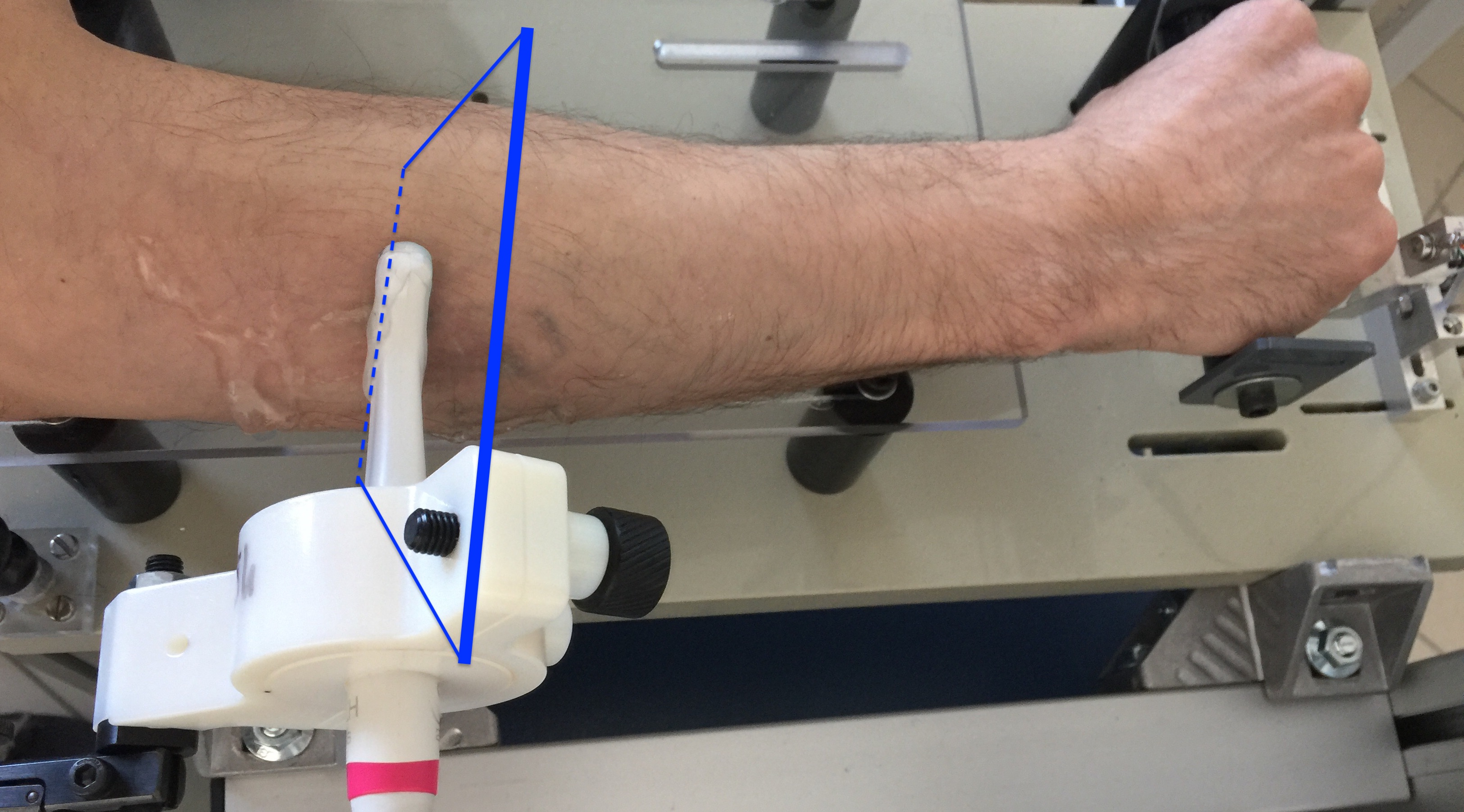}
        \includegraphics[width = 0.49\textwidth]{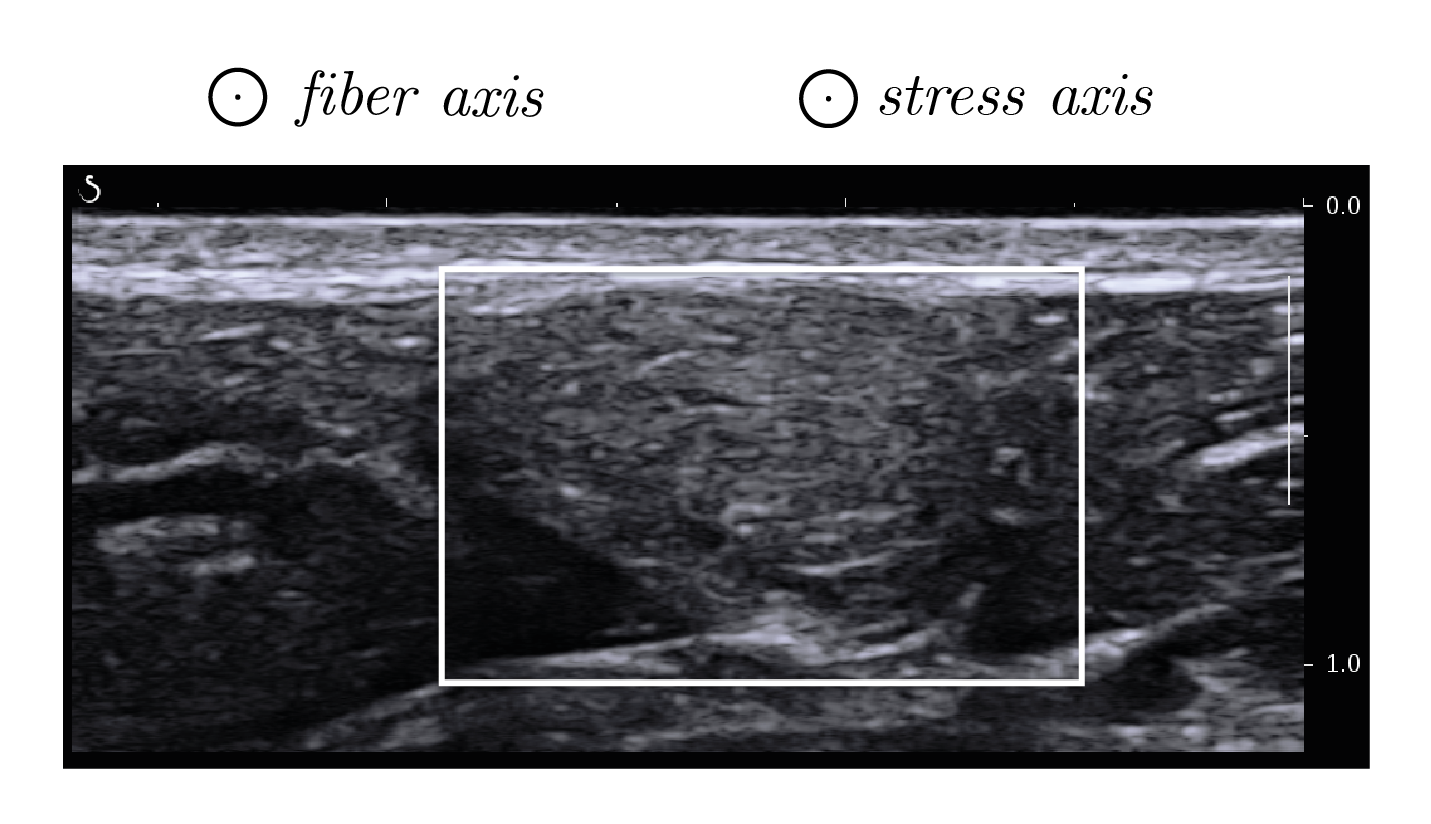}
   \caption{
     {\small
     (a) Probe placed above the right hand \textit{flexor digiti minimi} muscle perpendicularly to the fibre direction. (b) Corresponding Bmode image (26.7 $\times$ 14.7 mm), showing that the muscle  is almost circular, see center of the image.}}
     \label{BmodeTrans}
\end{figure}

%++++++++++++++++++++++++++++

\subsection{Muscle and imaging plane}

%+++++++++++++++++++++++++++++

The structure of muscles is complicated by inhomogeneities in fibre orientation and interfaces between  fibre bundles. 
To isolate the behaviour of an individual muscle, unaffected by the complementary or antagonistic actions of other muscles, we select the \textit{flexor digiti minimi} muscle, which extends the hand's little finger. 

This muscle is the only one involved in the little finger's extension, it has homogeneous fibre orientation (TI symmetry), and it is close to the epidermis, which matters for the relatively high frequency probe (SLH20-6 SSI probe, 12 MHz center frequency) used for the SWE measurements. 
Furthermore, as shown in Figures \ref{BmodeLongi}-\ref{BmodeTrans}, this muscle is convenient for probing, as it is situated on the side at the top of the forearm and spans over a distance longer than the 26.7 mm imaging width of the SLH20-6 probe. 

The quality of the ultrasound imaging system is important for the probe positioning and the  localisation of the muscle, which has a diameter of about 0.6 cm. 
One way to precisely localize the muscle on the image is to move slowly the little finger and look at the lateral tissue displacement in real time, with the probe positioned in the fibre direction. 
As shown on Figures \ref{BmodeLongi}(a) and \ref{BmodeTrans}(a), the probe is held axially and transversely to the fibre direction by a free arm stand, which can be locked in the desired position. 
We took particular care not to apply pressure with the probe on the skin surface. 
Bmode images of the right hand \textit{flexor digiti minimi} are shown on Figures \ref{BmodeLongi}(b) and \ref{BmodeTrans}(b), for a representative subject at rest. 
The image dimension is 14.7 mm depth by 26.7 mm width, giving an idea of the small size of the muscle. 
The white boxes delimit the region of interest, where SWE data are acquired. 

\begin{figure}%[H]
    \centering
    \begin{subfigure}[b]{0.48\textwidth}
        \centering
        \includegraphics[keepaspectratio=true,height = .3\textheight]{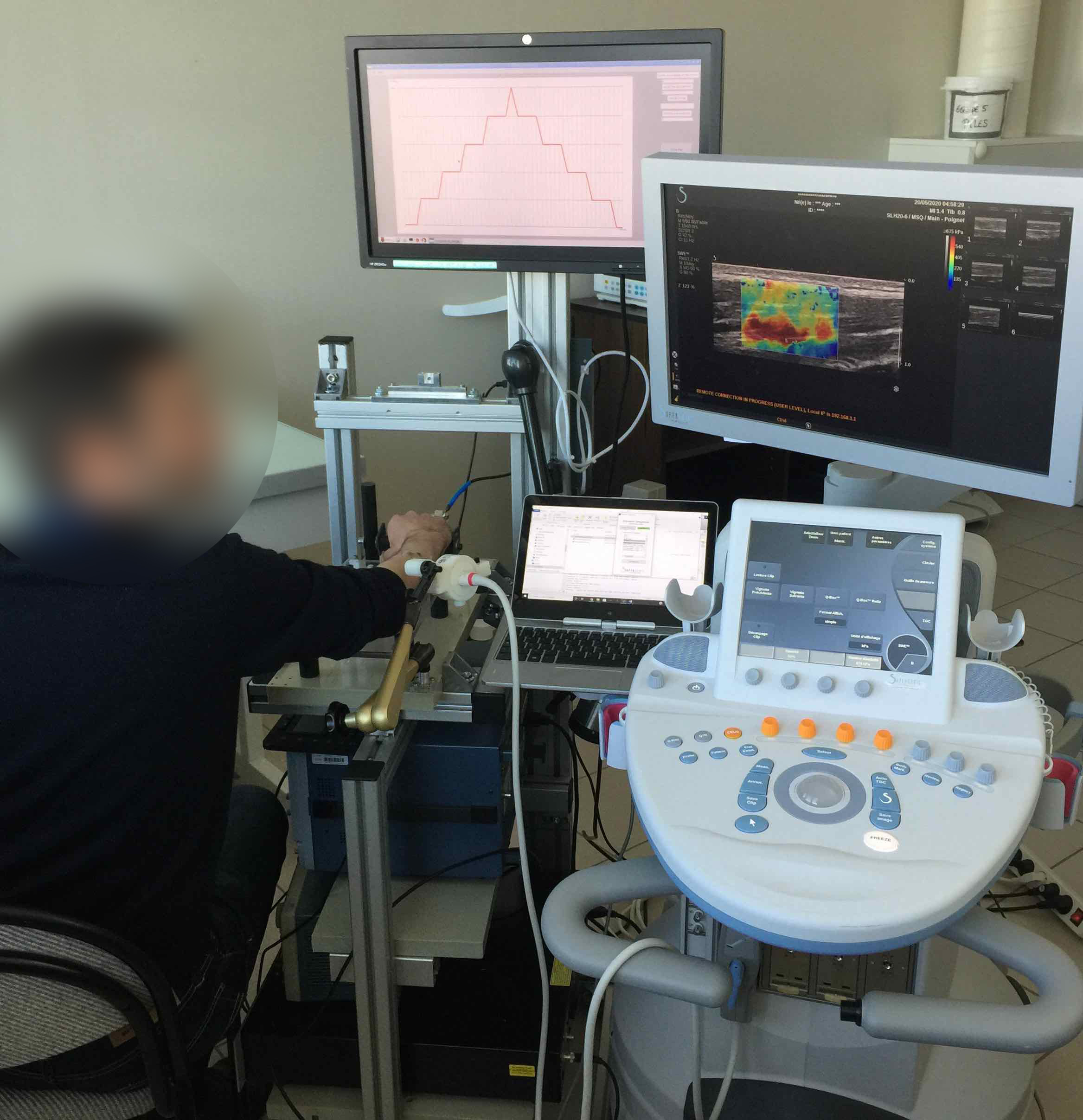}
        \caption{
          {\small
          The subject is seated with their right elbow flexed at $135^\circ$ ($180^\circ$ corresponds to the full extension of the elbow) and positioned vertically at approximately $70^\circ$ to the body.}} \label{ExpSetup}
    \end{subfigure}
    \quad
    \begin{subfigure}[b]{0.48\textwidth}
        \centering
        \includegraphics[keepaspectratio=true,height = .3\textheight]{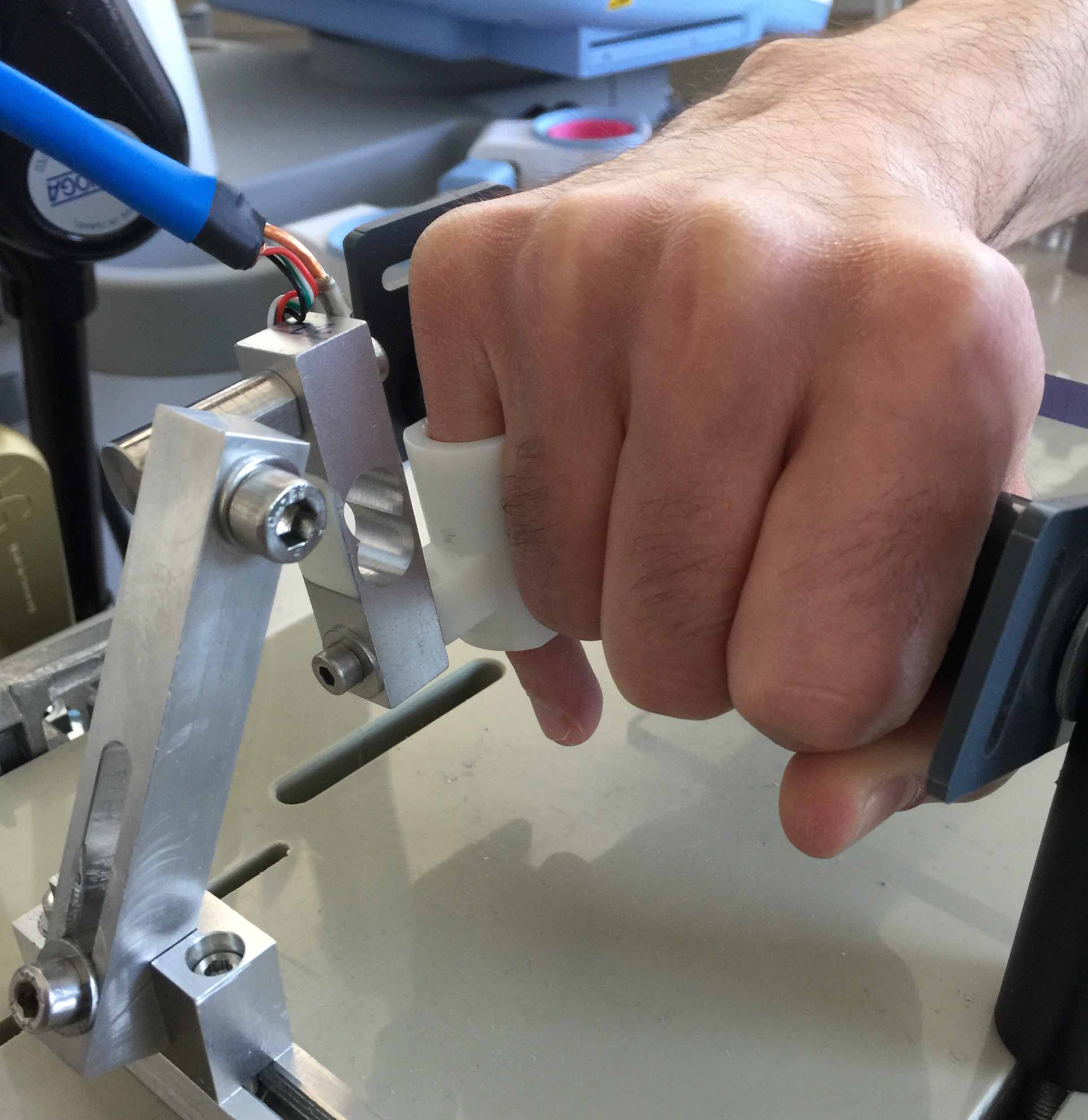}
        \caption{
          {\small
          The little finger's first phalanx is placed vertically and in contact with a cylindrical rigid interface. The finger is aligned with the force sensor.} }\label{ForceSensor}
     \end{subfigure}
     \caption{
     {\small
     Experimental setup including the custom-made force measurement system and the US Aixplorer imaging system running the SonicLab V12 research software to upload the Shear Wave Elastography (SWE) sequence.}
     } \label{ProtoForce}
\end{figure}

%+++++++++++++++++++++++++++++++++++++++++

\subsection{Protocol and participants} \label{Protocol}

%++++++++++++++++++++++++++++++++++++++++++

Figure \ref{ProtoForce} shows the experimental setup including the custom-made force measurement system and the ultrasound imaging system. 

Two women and four men, all right-handed, took part in this feasibility trial. 
They were informed of the possible risk and discomfort associated with the experimental procedures prior to giving their written consent to participate. 
Neither pregnant women nor persons under guardianship were included. 
The experimental design of this study was approved by the local Ethical Committee (Number ID RCB: 2020-A01601-38) and was carried out in accordance with the Declaration of Helsinki. 

The subjects are seated with their right elbow flexed to 135$^\circ$ (180$^\circ$ corresponds to the full extension of the elbow) and positioned vertically at approximately 70$^\circ$ to the body. 
The first phalanx of the little finger is placed vertically and in contact with a cylindrical rigid interface (Figure \ref{ForceSensor}), so that it is aligned with the calibrated force sensor  (micro load cell CZL635-20) and at rest. 
A lever arm, approximately 3 cm long, is placed between the force measuring point and the axis of rotation of the finger. That short distance might lead to a small difference between the magnitude of the force created inside the extensor muscle and the force measured by the sensor. At any rate, we assume that this effect is reflected by a small proportionality factor, which we neglect in our analysis. 
First, we asked the subjects to perform three maximum isometric voluntary contractions (MVC) lasting at least 3 seconds and separated by 30 seconds of recovery. 
The largest of the three forces was considered as the maximum voluntary force and was used to normalize subsequent submaximal contractions. 

Table \ref{table 1} details the age, genre, handedness, maximum voluntary force developed with the little finger, diameter of the muscle and finally, the maximum axial stress $\sigma_{\rm 11Max}$ calculated by dividing the MVC force by the current muscle cross-surface area (obtained using the Bmode image in the direction transverse to the fibre direction). 
Interestingly, in spite of their great age difference (40 years), Subject\#2 and Subject\#3 (both male) develop the same maximum force magnitude (the maximum lifted load difference is only 26 g) but Subject\#2 has twice the muscle cross-surface area as Subject\#3. 
Thus the maximum voluntary stress  $\sigma_{\rm 11Max}$ induced by Subject\#2 is half that of Subject\#3. 
On the other hand, the axial stress $\sigma_{\rm 11Max}$ obtained by Subject\#5 is one of the smallest in the cohort, while its maximum lifted load is the largest. 
We also note that the maximum voluntary force  range is large, from a low of 5.25 N for Subject\#4 to a high of 9.06 N for Subject\#5.
\begin{table}[h]
\caption{{\small Age, Gender, R/L Handedness, Maximum Lifted Load, Maximum Volontary Force, \textit{flexor digiti minimi} muscle surface and maximum axial stress $\sigma_{\rm 11Max}$ for the six healthy volunteers involved in the feasibility trial.}}
\footnotesize\rm
\begin{tabular}{@{}*{8}{l}}
\br
Subject & Age & Genre &  Handedness & Maximum & Maximum & Muscle & $\sigma_{\rm 11Max}$\\
&&&&  Load Lifted & Volontary Force & Surface &\\
&(years) & (M/F) & R/L& (g) &(N) &(cm$^2$) & (kPa)\\
\mr
\#1 &22 &M  &R  &630    &6.30   &0.28   &225\\
\#2 &62 &M  &R  &725    &7.25   &0.27   &268\\
\#3 &22 &F  &R  &699    &6.99   &0.14   &499\\
\#4 &25 &F  &R  &525    &5.25   &0.2    &262\\
\#5 &40 &M  &R  &906    &9.06   &0.5    &181\\
\#6 &32 &M  &R  &902    &9.02   &0.55   &164\\
\br
\end{tabular}\label{table 1}
\end{table}

Then the participants were asked to perform five voluntary contractions at levels corresponding to 4, 8, 12, 16, 20\% of MVC. They had to stay 4 seconds at each stage before moving to the next level. 
This period provided sufficient time to save the SWE image on the Aixplorer (and to allow for some viscous dissipation). 
To control the force steps, the participants followed a visual feedback displayed on a monitor placed in front of them, see Figure \ref{ExpSetup}.
It turned out to be difficult for some subjects to maintain precisely a constant force, especially at the  16\% and 20\% levels of the maximal voluntary contraction. 
For this reason we conducted our inverse analysis for measurements up to 12\% only, see results in Section  \ref{Results}.

%++++++++++++++++++++++++++++++++++++++++

\subsection{Shear Wave Elastography measurements}

%++++++++++++++++++++++++++++++++++++++++

We used the Shear Wave Elastography method to measure how shear elasticity changed with the force applied by the volunteers during the isometric contraction protocol.

The SWE experiment is based on two steps: the generation of the shear waves and the ultrafast imaging  of their propagation. In our experiments, the central frequency of the fast imaging scheme is 7.5 MHz and the image repetition frequency is set to 14 kHz, adapted to muscle stiffness.

\begin{figure}[hb!]
    \centering
          \includegraphics[width = \textwidth]{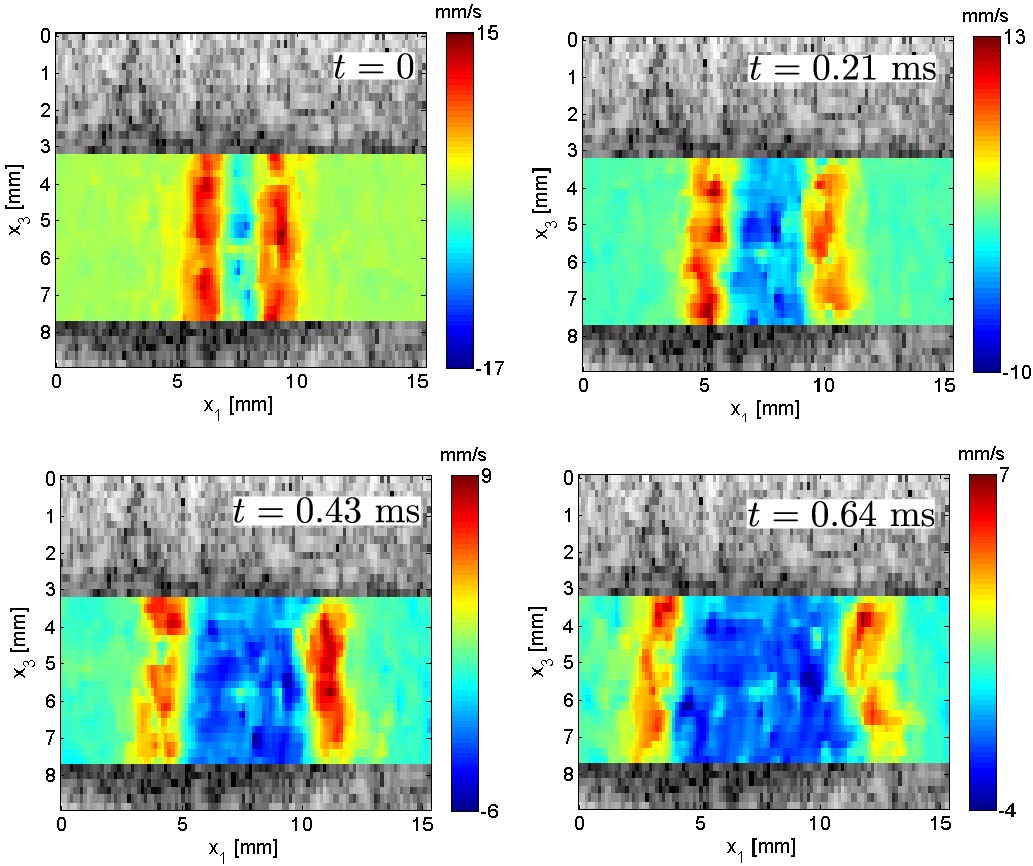}
      \caption{{\small Shear wave (SH mode) propagation along the \textit{flexor digiti minimi} muscle fibres for Subject\#5 at rest. 
     The wave propagates in the fibres direction $x_1$ and is polarized vertically along the $x_3$ axis.
     The color gives the tissue particle velocity at each location, see scale on the right (which adapts for better visualization).}}
     \label{PropaSWaveParaFibers}
\end{figure}

There are 48 frames in the temporal dimension, with a spatial resolution of 71.4 $\mu$s. Hence, we record the SW propagation for 3.4 ms, which is sufficient to follow the wave propagating along the width of the probe. The time $t=0$ ms in Figure \ref{PropaSWaveParaFibers} correspond to the beginning of the SWE data acquisition which is situated after the beginning of the push sequence.
There are 44 sampling points along the vertical axis, between the depths of 2.3 mm and 11.2 mm, with an axial resolution corresponding to one ultrasonic wavelength at 7.5 MHz, i.e 205 $\mu$m.
There are 110 sampling points along the $x_1$ axis, between $(x_1)_0 = $ 1.5 mm and $(x_1)_\text{Max} = $ 16.8 mm, with the lateral resolution of SLH20-6 probe pitch being 140 $\mu$m.

An SWE acquisition consists of five pushing lines positioned at -0.23, 3.77, 7.77, 11.77, 15.77 mm, with two focal points at 6.7 mm and 9.7 mm.

%%%%%%%%%%%%%%%%%%

\section{Results} \label{Results}

%%%%%%%%%%%%%%%%%%

%+++++++++++++++++++++++++++

\subsection{Propagation along the fibres}

%+++++++++++++++++++++++++++

Figure \ref{PropaSWaveParaFibers} shows the shear wave propagation induced in the muscle by the ultrasonic transient radiation force for Subject\#5 at rest. 
The radiation force is applied vertically along the positive $x_3$ axis. 
The wave propagates in the fibre direction along the $x_1$ axis and is polarized along the $x_3$ axis. 
The  propagation is presented at four different times: $t=0, 0.21, 0.43, 0.64$ ms, with a color scheme for the speed value, superimposed onto the Bmode image. 
The color scale is adapted for each image to take into account wave attenuation and enhance visualization.
Note that here the Bmode image is obtained from the shear wave tracking sequence and has a lower quality than the Bmode image shown in Figure \ref{BmodeLongi}.  
For this figure, we selected the third push zone situated at the lateral position 7.77 mm.

 \begin{figure}[hb!]
    \centering
        \includegraphics[width = \textwidth]{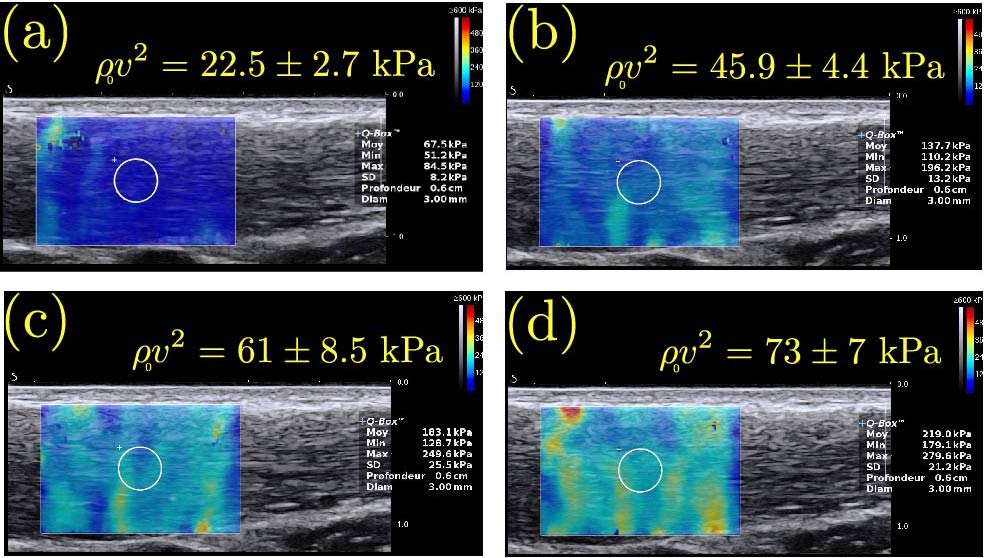}
  \caption{
  {\small 
  SWE analysis for  Subject\#5 at four levels of voluntary contraction along the \textit{flexor digiti minimi} muscle fibres. 
(a): at rest,  (b): 4\% MVC, (c): 8\% MVC, (d): 12\% MVC.
From the Aixplorer measure we deduce the average wave speed in the ROI (disc inside white circle) and compute the apparent axial shear modulus $\mu_\parallel(\sigma_{11}) = \rho_{\rm 0} v^2$. For this subject, it increases with the applied stress.
}
}
\label{Rho0v2ESigma11}
\end{figure}

At time $t=0.64$ ms, we can clearly see that the lower part of the shear wave front is ahead of the other parts of the wave front, indicating that the wave propagates faster in the muscle. 
Thus, we select the region of interest (ROI) in that part of the picture, from $x_3 = 6.2$ to $x_3=7.7$ mm (6 points), where the speed $v$ is assumed homogeneous to average the lateral propagation and improve the signal-to-noise ratio.

We assume that the phase speed dispersion is small at that the SWE measurement gives the speed of all shear waves with different frequencies in the wave packet.
Further, we assume that viscosity might attenuate the amplitude of the wave, but does not modify its speed noticeably \citep{bercoff2004role}.

Figures \ref{Rho0v2ESigma11} show the measurements given by the SWE diagnostic mode of the Aixplorer, obtained for Subject\#5 at four levels of voluntary contraction: 0, 4, 8, 12 \% of MVC, corresponding to $\sigma_{11}$ equal to 0, 7.2, 14.5, and 21.7 kPa, respectively. 
The machine gives a ``stiffness'' value, obtained by multiplying $\rho_{\rm 0} v^2$ by 3 to yield the apparent isotropic Young modulus. However here the material is anisotropic and we cannot use that formula. 
Instead we simply divide back  the machine mean value over the selected ROI (say $67.5\pm 8.2$ kPa at rest, Figure  \ref{Rho0v2ESigma11}(a)) by 3 (to obtain $\rho_{\rm 0} v^2 = 22.5\pm 2.7$ kPa at rest, for example).

\begin{figure}[hb!]
    \centering
        \includegraphics[width= 1\textwidth]{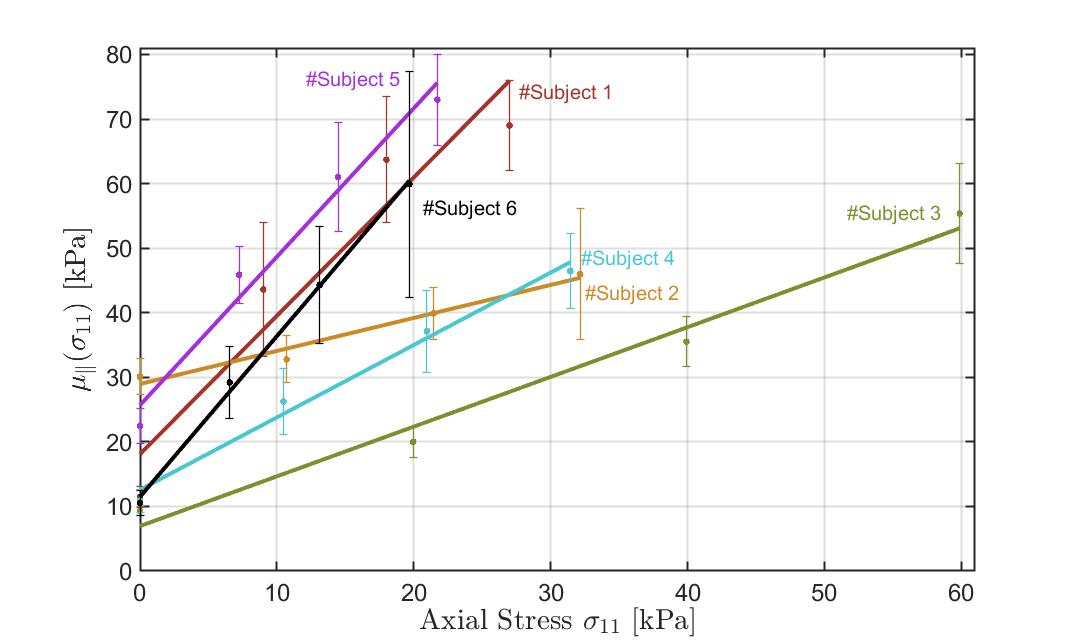}
        \includegraphics[width= 1\textwidth]{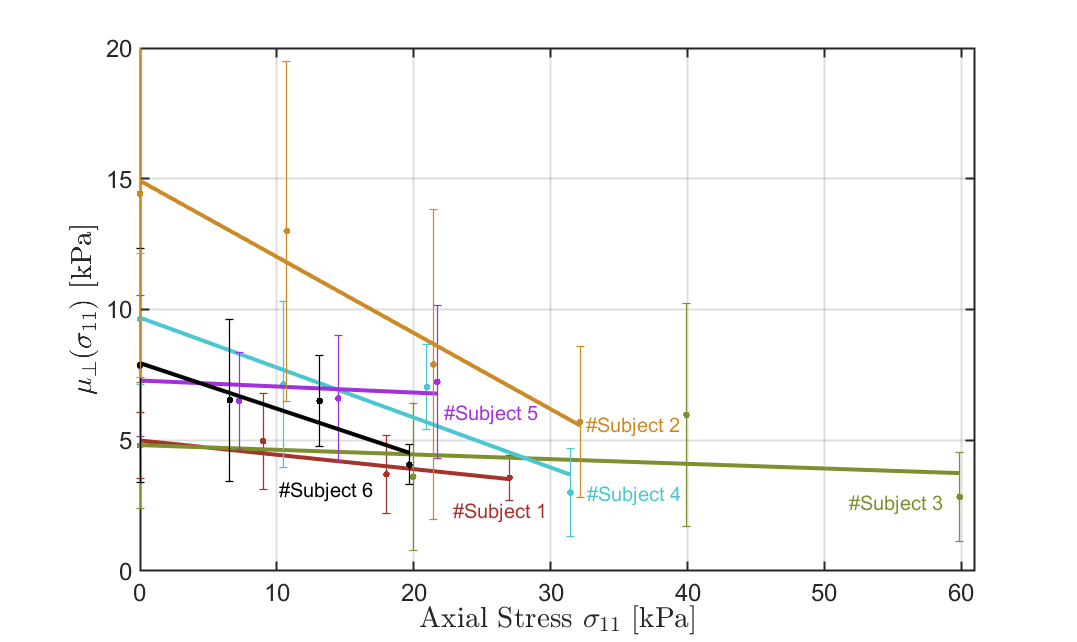}
         \caption{
         {\small
    Apparent shear elastic modulus $\rho_{\rm 0} v^2\left(\sigma_{11}\right)$ changes with the axial stress $\sigma_{\rm 11}$, measured \textit{in vivo} by SWE on the \textit{flexor digiti minimi} muscle (a) along the fibres axis $\mu_\parallel\left(\sigma_{11}\right)$ and (b) across the fibres axis $\mu_\bot\left(\sigma_{11}\right)$ for the six subjects of our feasibility study.}}
     \label{Mu_Subjects}
\end{figure}

We collected the measurements on Figure \ref{Mu_Subjects}(a) for the six volunteers. We notice a linear variation  of $\mu_{\parallel}\left(\sigma_{11} \right)$ with $\sigma_{11}$ in the fibres direction, similar to the behaviour obtained experimentally by \cite{Bouillard2011} on the \textit{abductor digitimi minimi} muscle. 
This variation is also in line with our theoretical analysis, according to which $\mu_{\parallel}\left(\sigma_{11}\right) = \rho_{\rm 0} v^2$ is given by (\ref{resultSH}) with $\theta=0$ as
 \begin{equation}
\mu_{\parallel}\left(\sigma_{11} \right)=\mu_{\rm L}-\beta_\parallel\sigma_{\rm 11},
\label{vbABeta13}
\end{equation}
where the non-dimensional coefficient of nonlinearity $\beta_\parallel$  is given by (\ref{BetaParr}).

For all six subjects, $\mu_{\parallel}\left(\sigma_{11}\right)$ increases with $\sigma_{\rm 11}$, so that $\beta_\parallel<0$ in the cohort. 

For the curve-fitting exercise  determining the quantities $\mu_{\rm L}$ and $\beta_\parallel$, we use the Matlab \emph{robustfit} algorithm which allocates lower weight to points that do not fit well. 
It also outputs the coefficient of determination $R^2$ and the root mean squared error RMS$_{\rm e}$. 

%+++++++++++++++++++++++++++

\subsection{Propagation across the fibres}

%+++++++++++++++++++++++++++

In the  direction transverse to the muscle fibres, the shear wave is highly scattered by heterogeneities, which induces a poor signal-to-noise ratio for frequency analysis \citep{Deffieux2008SWS}. 

\begin{figure}[hb!]
            \includegraphics[width=\textwidth]{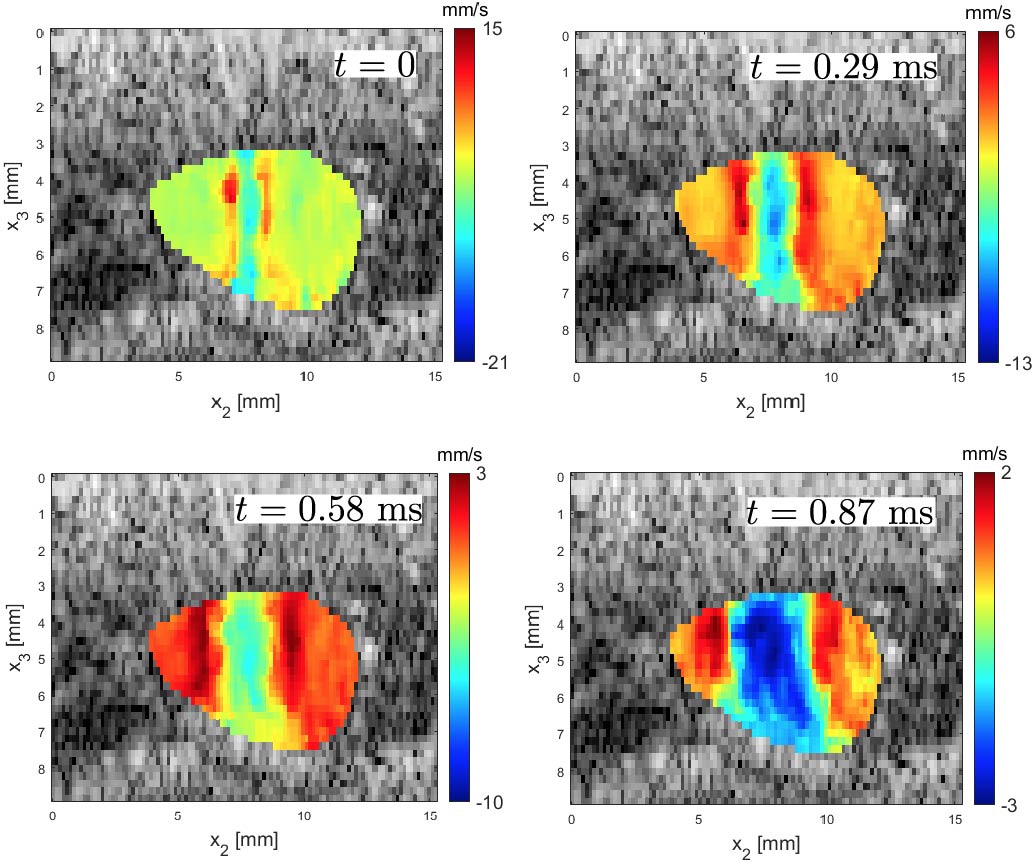}
           \caption{
     {\small
    Shear wave (SH mode) propagation transversely to the \textit{flexor digiti minimi} muscle fibres for Subject\#5 at rest. 
    The wave propagates along the $x_2$ axis and is polarized vertically along the $x_3$ axis. The color gives the tissue particle velocity at each location, see scale on the right (which adapts for better visualization)}}
     \label{PropaSWaveTransvFibers}
\end{figure}

Figures \ref{PropaSWaveTransvFibers} show the shear wave propagation perpendicularly to the fibres axis  for Subject \#5, at four different times: $t = 0, 0.29, 0.58, 0.87$ ms.
Superimposed onto the Bmode image, we show the  propagation inside the muscle only, which has a quasi-circular shape with a diameter of approximately 8 mm (see Figure \ref{BmodeTrans} for a more precise localisation of the muscle with a better Bmode image quality).

\begin{figure}%[H]
  \includegraphics[width=\textwidth]{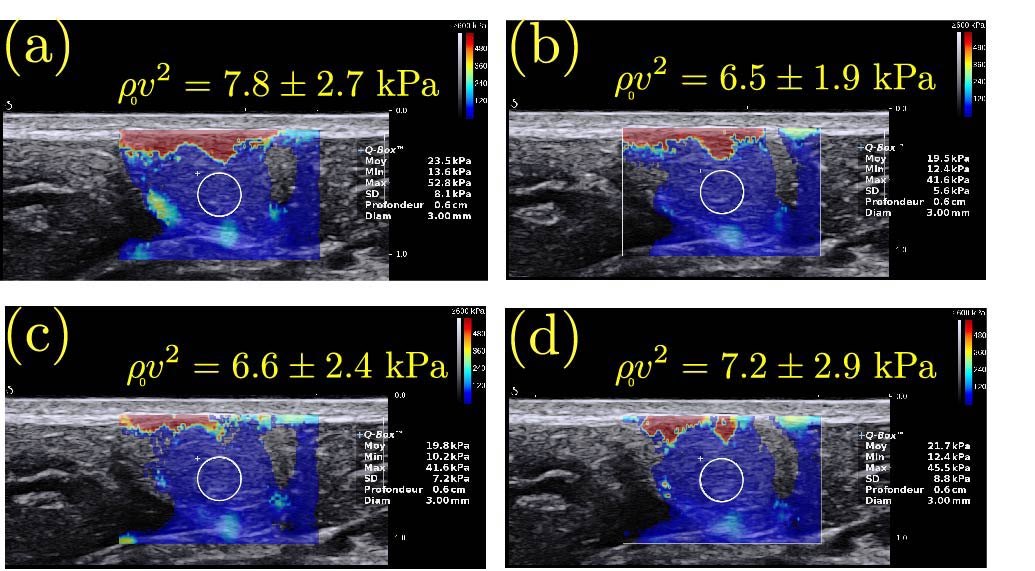}
  \caption{
  {\small
  SWE analysis for Subject\#5 at four levels of voluntary contraction transversely to the \textit{flexor digiti minimi} muscle fibres.
  (a): at rest,  (b): 4\% MVC, (c): 8\% MVC, (d): 12\% MVC.
  From the Aixplorer measure we deduce the average wave speed in the ROI (disc inside white circle) and compute the apparent transverse shear modulus $\mu_\perp(\sigma_{11}) =\rho_{\rm 0} v^2$. For this subject, $\mu_{\perp}$ is almost unchanged as $\sigma_{11}$ increases.
  }} \label{Rho0v2TSigma11}
\end{figure}

Figures \ref{Rho0v2TSigma11} show the values of the apparent shear modulus $\mu_\perp(\sigma_{11}) = \rho_{\rm 0} v^2$ in the transverse direction, for Subject\#5. Again, we present measurements up to 12\% of MVC. 

According to our theoretical analysis, $\mu_\bot\left(\sigma_{11}\right)$ is equal to $\rho_{\rm 0} v^2$ given by (\ref{resultSH}) with $\theta=90$° as
\begin{equation}
\mu_\bot\left(\sigma_{11}\right)=\mu_{\rm T}+\beta_{\perp}\sigma_{\rm 11},\label{resultatT}
\end{equation}

where the non-dimensional coefficient of nonlinearity $\beta_{\perp}$ is given by (\ref{BetaPerp}).

Again, we use the formula from acoustic-elasticity theory to produce a linear fit to the data.
In contrast to the case of propagation along the fibres, we find that $\mu_\perp\left(\sigma_{11}\right)$ does not increase with the axial stress $\sigma_{11}$, but decreases slightly for Subject\#5.
Other subjects lead to different behaviours, as can be checked on Figure \ref{Mu_Subjects}(b), but $\beta_{\perp}$ is always negative in the cohort.
We summarise the results in Table \ref{table 2}.
\begin{table}[h]
\caption{
  {\small
Axial shear elastic modulus at rest $\mu_{\rm L}$, axial nonlinearity coefficient  $\beta_{\parallel}$, transverse shear elastic modulus at rest $\mu_{\rm T}$, transverse nonlinearity parameter $\beta_{\perp}$, 12\% of maximum axial stress $\sigma_{\rm 11Max}$ for the \textit{flexor digiti minimi} muscle of the six healthy volunteers involved in the study.}
}
\footnotesize\rm
\begin{tabular}{@{}*{8}{l}}
\br
Subject & $\mu_{\rm L}$ & $\beta_{\parallel}$ & $R^2$  & $\mu_{\rm T}$ & $\beta_{\perp}$ & $R^2$ & 0.12$\sigma_{\rm 11Max}$\\
&(kPa)  &  &  & (kPa)  &  &   & (kPa)  \\
\mr
\#1 & 18.1 $\pm$ 8.9 & -2.14 $\pm$ 0.53 & 0.89 & 5.0 $\pm$ 0.4 & -0.05 $\pm$ 0.02 & 0.72 & 27.0\\
\#2 & 29.0 $\pm$ 1.3 & -0.51 $\pm$ 0.07 & 0.97 & 14.9 $\pm$ 1.0 & -0.29 $\pm$ 0.05 & 0.94 & 32.2\\
\#3 & 6.9 $\pm$ 3.0 & -0.77 $\pm$ 0.08 & 0.98 & 4.8 $\pm$ 1.5 & -0.02 $\pm$ 0.04 & 0.09 & 59.9\\
\#4 & 12.6 $\pm$ 2.1 & -1.12 $\pm$ 0.10 & 0.98 & 9.7 $\pm$ 1.0 & -0.19 $\pm$ 0.05 & 0.87 & 31.4\\
\#5 & 25.7 $\pm$ 3.9 & -2.30 $\pm$ 0.30  &  0.96 & 7.3 $\pm$ 0.7 & -0.02 $\pm$ 0.05 & 0.09 & 21.7\\
\#6 & 11.5 $\pm$ 1.2 & -2.50 $\pm$ 0.09 & 0.99 &  7.9 $\pm$ 0.6 & -0.17 $\pm$ 0.05 & 0.85 & 19.7\\
\br
\end{tabular}\label{table 2}
\end{table}

%%%%%%%%%%%%%

\section{Discussion}

%%%%%%%%%%%%%%

Using acousto-elasticity theory, we obtained  analytical expressions for the dependence of the SH shear wave speed as a function of the applied uniaxial stress in muscle, assuming that it behaves as a transversely isotropic, incompressible soft solid, and that the wave travels either along or transverse to the fibres.

For our experiments, we oriented the acoustic radiation force along the vertical axis and propagated the wave along or transverse to the \textit{flexor digiti minimi} muscle to avoid coupling of the shear horizontal (SH) mode with the shear vertical (SV) mode \citep{Rouze2020}.
We determined theoretically and experimentally the apparent shear elastic modulus $\mu\left(\sigma_{11}\right) = \rho_{\rm 0} v^2$, and found it varies linearly  with $\sigma_{11}$. 

We obtained analytical expressions for $\mu_\parallel\left(\sigma_{11}\right)=\mu_{\rm L}-\beta_{\parallel}\sigma_{11}$ in the fibre direction and for $\mu_\bot\left(\sigma_{11}\right)=\mu_{\rm T}+\beta_{\perp}\sigma_{11}$ transversely to the fibre direction. The coefficient $\beta_{\parallel}$ is a linear combination of the second-order elastic parameters $\mu_{\rm L}$, $\mu_{\rm T}$, $E_{\rm L}$, and the third-order moduli $A$, $\alpha_3$, $\alpha_5$; the coefficient $\beta_{\perp}$ is written in terms of only two second-order parameters $\mu_{\rm T}$, $E_{\rm L}$ and two third-order moduli $A$, $\alpha_3$. 
Neither coefficient involves the other third-order parameter $\alpha_4$. 

Our {\it in vivo} analysis of the six-volunteer cohort focused on the variation of the apparent shear elastic moduli with $\sigma_{11}$. 
The results show that these variations are very different across the cohort.

For the analysis of $\mu_\bot\left(\sigma_{11}\right)$, we distinguish three subgroups.

For Subjects \#2,4,6 (first group), we find $\beta_{\perp} = -0.29, -0.19, -0.17$  $(\pm0.05)$ , respectively, all negative, indicating that $\mu_\perp$ decreases with $\sigma_{11}$.
Here the connective tissue surrounding the muscle fibres softened under axial stress $\sigma_{11}$. 

For Subjects \#1,3,5, we find $\beta_{\perp} = -0.05  \pm 0.02, -0.02\pm 0.04, -0.02 \pm 0.05$, respectively, all small values, indicating that $\mu_{\perp}$ is almost unchanged as $\sigma_{11}$ increases. 
For these three subjects, the infinitesimal shear elastic moduli $\mu_\text{T}$ are almost the same: $\mu_{\rm T} = 5.0 \pm 0.4, 4.8 \pm 1.5, 7.3 \pm 0.7$ kPa, respectively. 
However, Subject\#3 has a much greater value of  12\% of maximum axial stress ($12\% \sigma_{\rm 11Max} = 59.9$ kPa) than Subjects\#1,5, who have approximately the same value (27.0, 21.7 kPa, respectively). 
Subject\#3 also has a much lower magnitude of coefficient $\beta_\parallel$ ($=-0.77\pm0.08$) than Subjects\#1,5 who have approximately the same $\beta_\parallel$ coefficient ($=-2.14\pm0.53, -2.3\pm0.30$). 
Thus, we separate Subjects\#1,5 (second group) from Subject\#3 (third group).

By comparing expressions (\ref{BetaParr}) and (\ref{BetaPerp}) for the $\beta$ nonlinearity coefficients, we see that only  the $\mu_{\rm L}$ and $\alpha_5$ parameters can explain why $\beta_{\parallel}$ is different between the second and the third group, because they do not appear in the expression (\ref{BetaPerp}) for $\beta_\perp$ (which is the same for these two groups). 
Hence we see that a higher value of $\mu_{\rm L}$ in (\ref{BetaParr}) results in a higher value of the coefficient $\beta_{\parallel}$ for subjects who have an identical $\beta_{\perp}$. 
This is indeed what we observed experimentally, see values in Table \ref{table 2}.

For the analysis of the axial apparent shear modulus $\mu_\parallel\left(\sigma_{11}\right)$, we also find three subgroups, according to the magnitude of the nonlinearity coefficient $\beta_\parallel$. Hence Subjects\#2,3 both present small magnitudes for $\beta_{\parallel}$, Subject\#4 present intermediate value, and Subjects\#1,5,6 all present high values (in that order).

For a contraction from rest to 12\% MVC, Subject\#6's apparent elastic modulus $\mu_\parallel\left(\sigma_{11}\right)$ increases by a factor 6, from 10.5 to 59.9 kPa, demonstrating his remarkable ability to recruit fibres to harden the muscle very quickly. 
On the \emph{peroneus longus} muscle of anaesthetised cats, \cite{Petit1990} found that the S motor units can produce high values of muscle stiffness, suggesting that for Subject\#6, the motor units ratio S/F might be very high. 
Subject\#6 also develops the smallest 12\% of maximum axial stress value of the cohort, consistent with a high S/F ratio because slow (S) motor units develop quite small tensions compared with fast fatigue-resistant (FR) and fast fatiguable (FF) units \cite{Petit1990}. 

For Subjects\#1,5,6 (third group), we note that the increase in the magnitude of $\beta_\parallel$ is associated with a decrease of the maximal axial stress 
$\sigma_\text{11Max}$. 

For Subjects\#2,3 (first group), we recorded  the smallest magnitude of the nonlinearity coefficient $\beta_{\parallel}$ ($-0.51\pm0.07, -0.77\pm0.08$, respectively). 
These two subjects develop respectively the second-highest (32.2 kPa) and the highest (59.9 kPa) maximum axial stress $12\% \sigma_{\rm 11Max}$ in the cohort. These results are consistent with a high presence of fast fibres (F) that do not harden rapidly the muscle, associated with a quite high tension \cite{Petit1990}. 

A natural follow-up on this study is to apply the method to patients with musculoskeletal disorders or neurodegenerative diseases. 
However, the method must be adapted because we found it was difficult for some healthy subjects to maintain a voluntary constant force during the time required  to measure the wave speed, a task which could prove even more challenging for patients. 
This adaptation could be achieved with a muscle contained hardening method, based on the corresponding nerve electro-stimulation associated with a synchronous measurement of force and elasticity. 
That set up would provide a calibrated and repeatable stimulation protocol as used for EMG measurements.
A better localisation of the muscle and a better probe positioning in relation to the fibres orientation could be achieved by linking in real time the ultrasound Bmode image with the corresponding slice plan of a pre-acquired MRI volume image. 
The effect of the distance between the force sensor and the muscle could also be quantified precisely.
Finally, the modelling can be improved, to include heterogeneity, viscosity and dispersion. 

%%%%%%%%%%%

\section{Conclusion}

%%%%%%%%%%

The quantification of the elastic nonlinearity of biological tissues can prove to be a most valuable tool for the early diagnosis of musculoskeletal disorders. 
Here, we developed an acousto-elasticity theory to study the propagation of small-amplitude plane body waves in deformed transversely isotropic incompressible solids. 
For the shear horizontal (SH) mode, we obtained a linear relation between the squared wave speed $\rho_{\rm 0} v^2$ and the applied axial stress $\sigma_{\rm 11}$ using the second- and third-order elastic constants. 
Then we used this theory to analyse experimental results on skeletal striated muscle. 

With a cohort of six healthy volunteers, we uncovered a great diversity for the nonlinear behaviour of the \textit{flexor digiti minimi} muscle, for  the apparent shear modulus $\mu_\parallel\left(\sigma_{11}\right)$ along the fibres as well as for the transverse apparent shear modulus $\mu_\bot\left(\sigma_{11}\right)$.
Hence $\mu_\bot$ can decrease with $\sigma_{11}$ (Subjects\#2,4,6) or remain almost constant (Subjects\#1,3,5).
Meanwhile, $\mu_\parallel$ always increases with $\sigma_{11}$, and the rate of increase is highly correlated to the weakness of the maximum voluntary contraction produced by the volunteer. 

%%%%%%%%%%%%%%%

\ack{
We thank V\'eronique Marchand-Pauvert (Laboratoire d'Imagerie Biom\'edicale, Sorbonne Universit\'e) for helpful discussions about the role of motor units in muscle function and their relationship to muscle elasticity. 
We also thank Jean-Marc Gregoire and Jean-Yves Tartu (iBrain Laboratory, Universit\'e de Tours) for the development of the electronics and the mechanics of the force measurement device, respectively. This work is supported by the Institut du Biom\'edicament and by the INCA Plan Cancer BPALP project.}

%%%%%%%%%%%%%%%%

\newcommand{\newblock}{\ }

%%%%%%%%%%%%%%%%%%

%%%%%%%%%% Merge with supplemental materials %%%%%%%%%%
\pagebreak

\setcounter{equation}{0}
\setcounter{section}{0}
\setcounter{figure}{0}
\setcounter{table}{0}
\setcounter{page}{1}
\makeatletter
\renewcommand{\theequation}{S\arabic{equation}}
\renewcommand{\thefigure}{S\arabic{figure}}

\begin{center}
\textbf{\large Supplementary File:} \\ 
\textbf{Analytical calculations derived for the paper}\\
\textbf{ ``Acousto-elasticity of Transversely Isotropic Incompressible Soft Tissues: Characterization of Skeletal Striated Muscle"}
\end{center}
%%%%%%%%%% Merge with supplemental materials %%%%%%%%%%
%%%%%%%%%% Prefix a "S" to all equations, figures, tables and reset the counter %%%%%%%%%%
\vspace{2pt}
Jean-Pierre Remeni\'eras$^1$, Michel Destrade$^2$
\\[2pt]$^1$ UMR 1253, iBrain, Universit\'e de Tours, Inserm, Tours, France.
\\{$^2$ School of Mathematics, Statistics and Applied Mathematics, NUI Galway, University Road, Galway, Ireland.}
%
%\ead{jean-pierre.remenieras@univ-tours.fr}
%%\date{\today}
%
%\begin{abstract}
\\[12pt]
{\small
\textbf{Abstract.} This supplementary file details the analytical calculations of the acousto-elasticity method used in the paper ``Acousto-elasticity of Transversely Isotropic Incompressible Soft Tissues: Characterization of Skeletal Striated Muscle". 
For generality, we treat both the shear-horizontal (SH) and the shear-vertical (SV) propagation modes in homogeneous, transversely isotropic, incompressible solids subject to a uniaxial stress along the fibres. 
In the main paper, only the results for the (SH) mode are exploited, because  our experiments are only sensitive to this  polarisation.

Acousto-elasticity requires an expansion of the strain-energy density up to at least the third order in the strain.
Here we express the speed of the shear waves as a function of the second- and third-order elastic moduli and of the  propagation angle $\theta$ between the direction of the fibres and the direction of propagation. 
In the main paper, we take $\theta=0^\circ$ and $\theta=90^\circ$, in line with the experiments, but with the expressions calculated in this supplementary file, it is possible to perform a propagation analysis for any angle $\theta$, by rotating the probe or by using a 3D Shear Wave Elasticity method.
}
%\end{abstract}
%
%%\date{today}
%\vspace{2pc}
%\noindent{\it Keywords}: Acousto-elasticity theory, Transversely Isotropic soft solid, Third-order elastic constants, Incompressible TI material, strain-energy density.
%
%%\submitto{\PMB}
%\maketitle

%%%%%%%%%%%%%%%%%%%%%%%%%%%%%%%%%%%%%%%%%%%

\section{Uniaxial stress in incompressible transversely isotropic solids}
\label{TI theory}

%%%%%%%%%%%%%%%%%%%%%%%%%%%%%%%%%%%%%%%%%%%

\begin{figure}[ht]
\includegraphics[width=.7\linewidth]{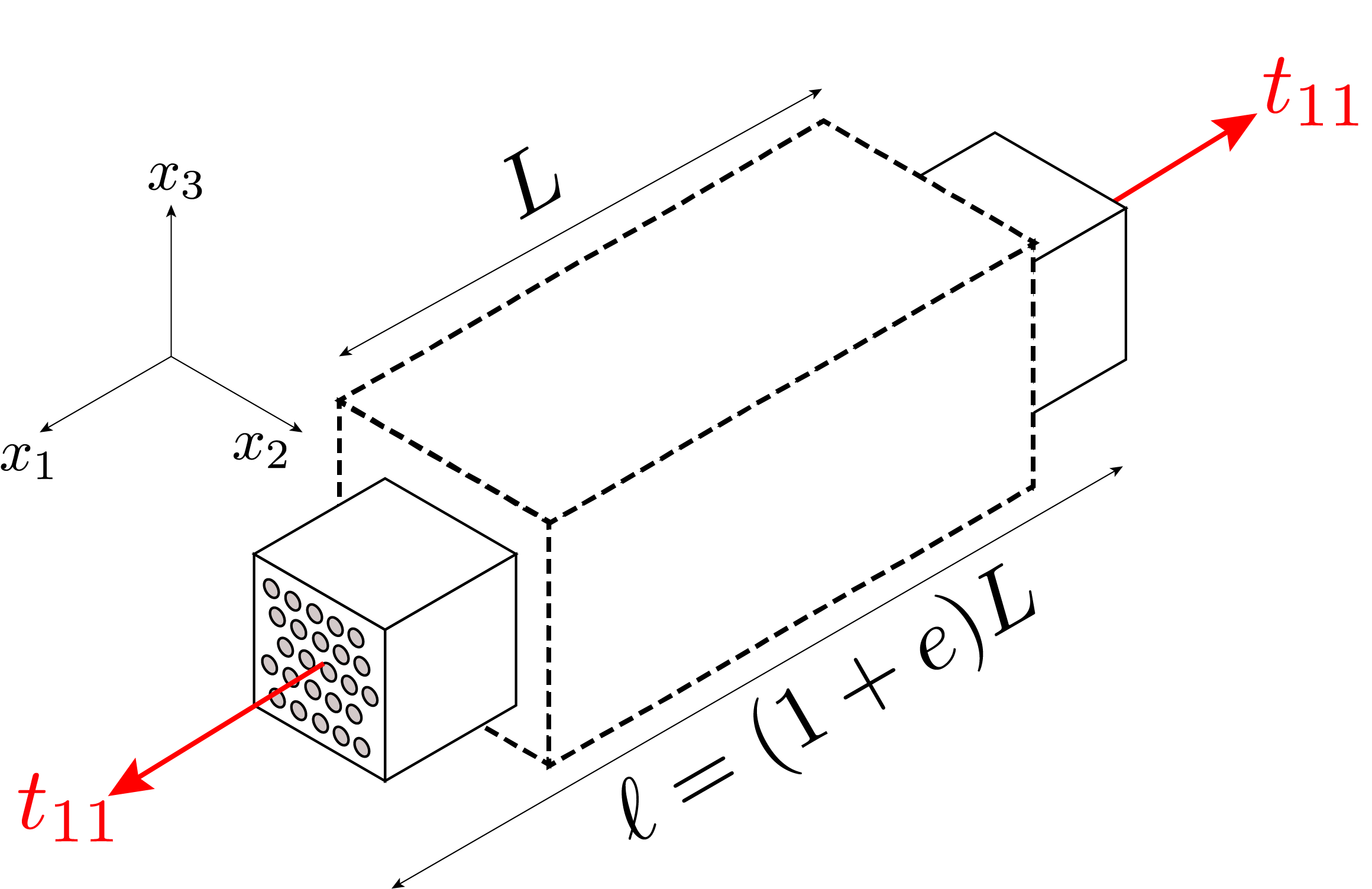}  
\centering
\caption{In our experiments, the volunteers apply a uniaxial stress of magnitude $\sigma_{\rm 11}$ along the direction of the fibres (the $x_{\rm 1}$ axis) during  voluntary contractions. In reaction, their muscle experiences the uniaxial Cauchy stress $t_{11} = - \sigma_{11}$, and its length is consequently changed by amount $e$ along $x_1$, and by amounts $-e/2$ along $x_2$ and $x_3$ (by symmetry and by incompressibility).}
\label{Uniaxial_Stress}
\end{figure}

We model muscles as soft incompressible solids with one preferred direction, associated with a family of parallel fibres, see Figure \ref{Uniaxial_Stress} for a description of the kinematics and physics of the model.

Transversely isotropic (TI), linearly elastic, \emph{compressible} solids are described by five independent constants, for example the following set \citep{Rouze2020}: $\mu_{\rm L},E_{\rm L}, E_{\rm T}, \nu_{\rm TT},\nu_{\rm LT}$, 
where $ \mu_{\rm L}$ is the shear elastic modulus relative to deformations along the fibres, $E_{\rm L}$, $E_{\rm T}$ are the Young moduli along, and transverse to, the fibres, respectively, and $\nu_{\rm TT}$, $\nu_{\rm LT}$ are the Poisson ratios in these directions. 
The shear elastic modulus $\mu_{\rm T}$ relative to the transverse direction is
\begin{equation} 
\mu_{\rm T}=\frac{ E_{\rm T}}{2\left(1+ \nu_{\rm TT}\right)}. 
\label{eq1} 
\end{equation}
For \emph{incompressible} TI materials, there is no volume change. 
This constraint leads to the following relations (see \cite{Rouze2020} for details),
\begin{equation} 
\nu_{\rm LT} = \textstyle \frac{1}{2}, \qquad \nu_{\rm TT}=1-\frac{E_{\rm T}}{2E_{\rm L}}. \label{eq2} 
\end{equation}
Thus, only three independent constants are required to fully describe a given transversely isotropic, linearly elastic, incompressible solid. 
Here we choose the three material parameters $\mu_{\rm T}$, $\mu_{\rm L}$, and $E_{\rm L}$, as proposed by \cite{Li2016}. 
Other, equivalent choices can be made \citep{Chadwick1993,Rouze2013,Papazoglou2006}, for instance by using $\mu_{\rm T}$, $\mu_{\rm L}$, and $\frac{E_{\rm L}}{E_{\rm T}}$. By inserting $\nu_{\rm TT}$ given by (\ref{eq2}) into (\ref{eq1}) we obtain \begin{equation} 
E_{\rm L}=\left(4\frac{E_{\rm L}}{E_{\rm T}}-1\right)\mu_{\rm T}, \label{eq3} 
\end{equation}
which makes the link between the two descriptions. 

As shown in Figure \ref{Uniaxial_Stress}, we call $x_{\rm 1}$ the axis along the fibres and $t_{\rm 11}$ the uniaxial Cauchy stress experienced by the muscle in reaction to the stress $\sigma_{11}$ applied by the volunteers in that direction during the voluntary contractions. 
The resulting extension in that direction is $e$ ($e>0$: elongation, $e<0$: contraction).
Then, following  \cite{Chadwick1993}, we have
\begin{equation}
 t_{\rm 11} = -p+2\left(2\frac{E_{\rm L}}{E_{\rm T}}-1\right)\mu_{\rm T}e,
 \label{eq6} 
\end{equation}
where $p$ is a Lagrange multiplier introduced by the constraint of incompressibility (to be determined from initial/boundary conditions).
Here the lateral stresses are $t_{\rm 22} = t_{\rm 33}=0$ so that  
\begin{eqnarray}
 0 = -p+2\mu_{\rm T}\left(-\frac{e}{2}\right),
 \label{eq7} 
\end{eqnarray}
because the lateral extension is $-e/2$ by symmetry and incompressibility.
This equation yields $p$ and then,
\begin{eqnarray}
 t_{\rm 11} = \left(4\frac{E_{\rm L}}{E_{\rm T}}-1\right)\mu_{\rm T}E_{\rm 11},
 \label{eq9} 
\end{eqnarray}
which can be simplified using (\ref{eq3}) into the classical expression $t_{11} = E_{\rm L}e$. 
In terms of the uni-axial stress $\sigma_{11} = -t_{11}$ applied by the volunteers on the muscle, we have
\begin{eqnarray}
\sigma_{\rm 11} = - E_{\rm L}e.
\label{eq10} 
\end{eqnarray}
This relation  will be used to go from $(v-e)$ to $(v-\sigma_{11})$ formulations of the acoustic-elastic equations.

%%%%%%%%%%%%%%%%%%%%%%%%%%%%

\section{\label{Background} Third-order expansion of the strain energy in TI incompressible solids}

%%%%%%%%%%%%%%%%%%%%%%%%%%%%%%

Acousto-elasticity calls for a third-order  expansion of the elastic strain energy density $W$ in the powers of $\bm{E}$, the Green-Lagrange strain tensor. 

For transversely isotropic incompressible solids, the expansion can be written as \citep{Destrade2010second},
\begin{equation}
 W = \mu_{\text  T}I_{\rm2} + \alpha_1 I_{\rm4}^2+\alpha_2I_{\rm5} +\frac{A}{3}I_{\rm3}+\alpha_3I_{\rm2}I_{\rm4}+\alpha_4I_{\rm4}^3+\alpha_5I_{\rm4}I_{\rm5},
\label{W} 
\end{equation}
where the second-order elastic constants $\alpha_1$, $\alpha_2$ are given by 
\begin{equation}
\alpha_1=\textstyle\frac{1}{2}\left(E_{\rm L}+\mu_{\rm T}-4\mu_{\rm L}\right),
\qquad \alpha_2=2\left(\mu_{\rm L}-\mu_{\rm T}\right),
\label{ConstOrdr2} 
\end{equation}
and $A$, $\alpha_3$, $\alpha_4$ and $\alpha_5$ are third-order elastic constants. 
The strain invariants used in (\ref{W}) are
\begin{equation}
I_2 = \tr(\bm{E}^2), \quad
I_3 = \tr(\bm{E}^3), \quad
I_4 = \bm{A \cdot EA}, \quad
I_5 = \bm{A \cdot E}^2 \bm{A},
\label{InvariantI} 
\end{equation}
where $\vec{A}$ is the unit vector in the fibres direction when the solid is unloaded.

\begin{figure}[ht]
\includegraphics[width=.95\linewidth]{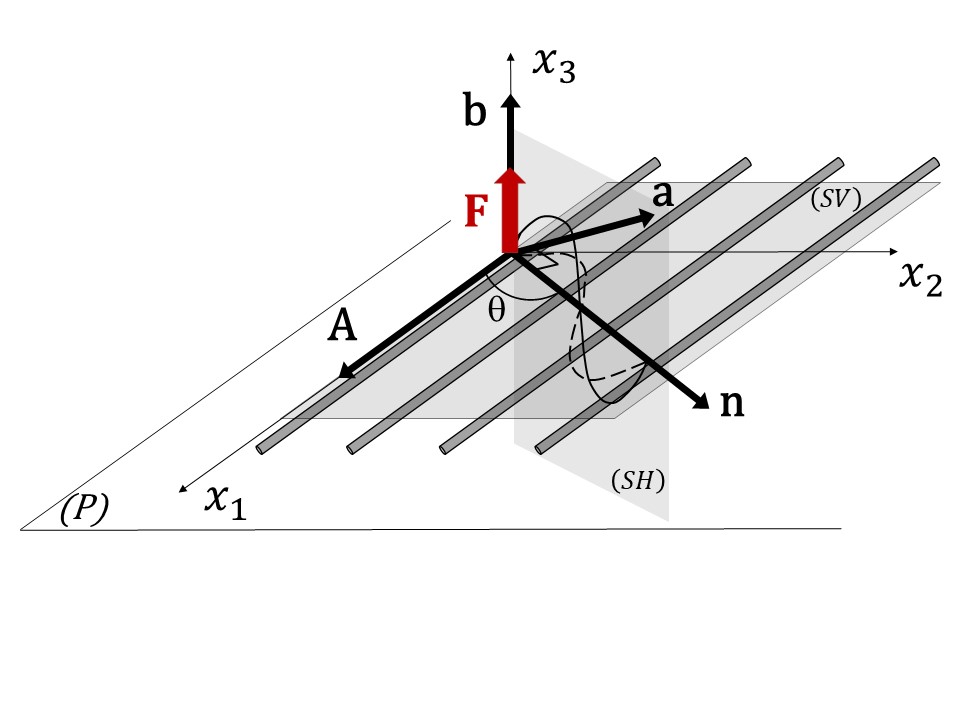}  
\centering
\caption{
We consider an incompressible transversely isotropic solid under uniaxial stress. Here $(P)$ is the ($\bm{A},\bm{n}$)-plane where $\bm{A}$ is a unit vector in the fibers direction when the solid is unloaded and at rest, and $\bm{n}$ is a unit vector in the direction of propagation. 
The  uni-axial tensile stress $\sigma_{11}$ is applied along the fibres. Two purely transverse waves propagate: the shear-vertical (SV) mode with  polarization $\bm{a}$ in the $(\bm{A},\bm{n})-$plane, and the shear-horizontal (SH) mode with polarization $\bm{b}$ normal to the $(P)-$plane. 
We call $\theta$  the angle between $\bm{n}$ and $\bm{A}$. In our experiments, the radiation force $\bm{F}$ is applied along the $x_{\rm 3}$ axis, and we measure the speed of waves travelling along the fibres ($\theta=0^\circ$) and transverse to the fibres ($\theta=90^\circ$). Ultrasound tracking measures the $x_{\rm 3}$ component of the shear wave displacement and is sensitive only to the (SH) propagation mode.}
\label{fig:Swaves}
\end{figure}

%%%%%%%%%%%%%%%%%%%%%%%

\section{\label{actotho} Small-amplitude plane body waves in the deformed soft tissue}

%%%%%%%%%%%%%%%%%%%%%%%%%

We now consider the propagation of small-amplitude plane body waves in a deformed soft tissue. 
 \cite{Destrade2010second} and \cite{Ogden2011}  show that investigating homogeneous plane wave propagation in TI incompressible solids is equivalent to solving the eigenproblem 
\begin{eqnarray}
\bm{\overline Q} \left(\bm{n}\right) \bm{a}=\rho_{\rm 0}v^2 \bm{a},
\label{EqOndeQ} 
\end{eqnarray}
where $\rho_0$ is the (constant) mass density, $\bm{a}$ is the unit vector along the direction of polarisation, $\overline{\bm Q}$ is the following symmetric tensor
\begin{equation}
\overline{\bm{Q}}\left(\bm{n}\right) =\left(\bm{I} -\bm{n}\otimes \bm{n} \right)\bm{Q}\left(\bm{n}\right)
\left(\bm{I}-\bm{n}\otimes \bm{n} \right), 
\label{Qbar} 
\end{equation}
with $\bm Q(\bm n)$ the (symmetric) acoustical tensor.
It is defined as
\begin{equation}
\left[\bm{Q}\left(\bm{n}\right)\right]_{ij} =\mathcal{A}_{{\rm0}piqj}n_pn_q,
\end{equation}
where
$\bm{\mathcal{A}_0}$ is the fourth-order tensor of instantaneous elastic moduli, with components \citep{Destrade2010second},
\begin{eqnarray}
\mathcal{A}_{{\rm0}piqj}=F_{p\alpha}F_{q\beta}\delta_{ij}\frac{\partial W}{\partial E_{\alpha\beta}}+F_{p\alpha}F_{q\beta}F_{j\nu}F_{i\gamma}\frac{\partial^2 W}{\partial E_{\alpha\gamma}\partial E_{\beta\nu}}.
\label{TensorA} 
\end{eqnarray}
Here $W$ is given by (\ref{W}) and $\bm{F}$ is the deformation gradient. 

In our case, the fibres are aligned with the direction of uniaxial stress and elongation, which is along the $x_1$-axis in the Eulerian description. 
Hence, at first-order in $e$, $\bm F = \text{Diag}\left(1+ e, 1 - e/2, 1- e/2\right)$, $\bm{E} = \text{Diag}\left(e,-e/2,-e/2\right)$, $I_4 = e$, $I_5 = 0$.

Because $\overline{\bm{Q}}\left(\bm{n}\right)$ is symmetric, its eigenvectors are orthogonal. 
By inspection we see that one eigenvector is $\bm{n}$, with eigenvalue $\rho_{\rm 0}v^2 =0$ indicating that no longitudinal wave may propagate in perfectly incompressible solids. The other two eigenvectors are $\bm{b}=\bm{A}\times \bm{n}$, along $x_3$, and $\bm{a} = \bm{b}\times \bm{n}$ which lies in the (SV) plane.
The corresponding  two shear velocities are given by
\begin{equation}
\rho_{\rm 0}v_\text{b}^2 = \mathcal{A}_{{\rm0}piqj}n_pn_q b_i b_j, \qquad 
\rho_{\rm 0}v_\text{a}^2 = \mathcal{A}_{{\rm0}piqj}n_pn_qa_ia_j.
\label{WaveEq} 
\end{equation} 

%%%%%%%%%%%%%%%%%%%%%%%%

\section{Acousto-elasticity of the (SH) wave}

%%%%%%%%%%%%%%%%%%%%%%%%%
 
The (SH) wave propagates along $\bm{n}=\left(\cos \theta, \sin \theta,0\right)$ and is polarised along $\bm{b}=\left(0,0,1\right)$, see Figure \ref{fig:Swaves}.

Using (\ref{TensorA}) and (\ref{WaveEq}), we find that the wave speed $v_{\rm b}$ is given by 
\begin{equation}
\rho_{\rm 0}v_{\rm b}^2 = \gamma_{13}\cos^2\theta + \gamma_{23}\sin^2\theta,\label{rhovb2}
\end{equation}
where
\begin{eqnarray}
& \gamma_{13} = \mathcal{A}_{01313} = F_{\rm11}^2\frac{\partial W}{\partial E_{\rm11}} +F_{\rm11}^2F_{\rm33}^2\frac{\partial^2 W}{\partial E_{\rm13}^2}, \nonumber \\[8pt]
& \gamma_{23} = \mathcal{A}_{\rm02323}=F_{\rm22}^2\frac{\partial W}{\partial E_{\rm22}} +F_{\rm22}^2F_{\rm33}^2\frac{\partial^2 W}{\partial E_{\rm23}^2}.
\end{eqnarray}

We first calculate $\gamma_{\rm13}$, using the strain energy $W$ in (\ref{W}) and the derivatives formulas of \cite{Destrade2010second}. 
We find, at the first order in $e$, that
\begin{eqnarray} 
& \frac{\partial I_2}{\partial E_{11}} = 2e, \qquad
\frac{\partial \left(I_4^2\right)}{\partial E_{11}} = 2e, \qquad
\frac{\partial I_{5}}{\partial E_{11}} = 2e, \nonumber \\[8pt]
& \frac{\partial I_{3}}{\partial E_{11}} = 0, \qquad
\frac{\partial \left(I_{2} I_{4}\right)}{\partial E_{11}} = 0, \qquad
\frac{\partial \left(I_{4}^3\right)}{\partial E_{11}} = 0,
\label{A1}
\end{eqnarray}
so that 
\begin{eqnarray}
F_{\rm11}^2\frac{\partial W}{\partial E_{\rm11}}=2\left(\mu_T+\alpha_1+\alpha_2\right)e.
\label{A2}\end{eqnarray}
The second term in the expression of $\gamma_{13}$ involves second derivatives of $W$. 
We obtain
\begin{eqnarray} 
& \frac{\partial^2 I_{\rm 2}}{\partial E_{\rm13}^2}=1, \qquad 
\frac{\partial^2 \left(I_{\rm 4}^2\right)}{\partial E_{\rm13}^2} = 0, \qquad
\frac{\partial^2 I_{\rm 5}}{\partial E_{\rm13}^2}={\frac{1}{2}},\qquad 
 \frac{\partial^2 I_{\rm 3}}{\partial E_{\rm13}^2} = {\frac{3}{4}e},  \nonumber\\[8pt]
& \frac{\partial^2\left(I_{\rm 2}I_{\rm 4}\right)}{\partial E_{\rm13}^2} =  e,\qquad
\frac{\partial^2 \left(I_{\rm 4}^3\right)}{\partial E_{\rm13}^2} = 0,\qquad
 \frac{\partial^2\left(I_{\rm 4}I_{\rm 5}\right)}{\partial E_{\rm13}^2} = {\frac{1}{2}e},
\label{A3}
\end{eqnarray}
so that
\begin{eqnarray}
F_{\rm11}^2F_{\rm33}^2\frac{\partial^2 W}{\partial E_{\rm13}^2}=\mu_T+\frac{\alpha_2}{2}+\left(\mu_T+\frac{\alpha_2}{2}+\frac{A}{4}+\alpha_3+\frac{\alpha_5}{2} \right)e.
\label{A4}\end{eqnarray}
Finally,  adding the two expressions, we obtain
\begin{eqnarray}
\gamma_{\rm13}  = \mu_T+\frac{\alpha_2}{2}+\left(3\mu_T+\frac{A}{4}+2\alpha_1+\frac{5}{2}\alpha_2+\alpha_3+\frac{\alpha_5}{2}\right)e.
\end{eqnarray} 
In terms of the classic linear moduli, see (\ref{ConstOrdr2}),  we have 
\begin{eqnarray}
\gamma_{\rm13} & = \mu_{\rm L}+\left(E_{\rm L}+\mu_{\rm L}-\mu_{\rm T}+\frac{A}{4}+\alpha_3+\frac{\alpha_5}{2}\right)e \nonumber \\
& =\mu_{\rm L}-\left[1+\frac{1}{E_{\rm L}}\left(\mu_{\rm L}-\mu_{\rm T}+\frac{A}{4}+\alpha_3+\frac{\alpha_5}{2}\right)\right]\sigma_{\rm 11},
\label{gamma13e}
\end{eqnarray}
where we used (\ref{eq10}) for the latter equality.

Now we compute $\gamma_{\rm23}=\mathcal{A}_{\rm02323}$.
We find in turn that
\begin{eqnarray} 
\frac{\partial I_{\rm 2}}{\partial E_{\rm22}}=-e, \qquad 
\frac{\partial \left(I_{\rm 4}^2\right)}{\partial E_{\rm22}} = \frac{\partial I_{\rm 5}}{\partial E_{\rm22}} = 
\frac{\partial I_{\rm 3}}{\partial E_{\rm22}} =  \frac{\partial \left(I_{\rm 2}I_{\rm 4}\right)}{\partial E_{\rm22}} = \frac{\partial \left(I_{\rm 4}^3\right)}{\partial E_{\rm22}}=0,
\label{A5}\end{eqnarray}
and that
\begin{eqnarray} 
& \frac{\partial^2 I_{\rm 2}}{\partial E_{\rm23}^2}=1, \qquad 
\frac{\partial^2 \left(I_{\rm 4}^2\right)}{\partial E_{\rm23}^2}=0, \qquad
\frac{\partial^2 I_{\rm 5}}{\partial E_{\rm23}^2}=0,  \qquad
\frac{\partial^2 I_{\rm 3}}{\partial E_{\rm23}^2} = -\frac{3}{2}e,  \nonumber\\[8pt]
& \frac{\partial^2\left(I_{\rm 2}I_{\rm 4}\right)}{\partial E_{\rm23}^2}=e, \qquad
\frac{\partial^2 \left(I_{\rm 4}^3\right)}{\partial E_{\rm23}^2}=0, \qquad 
\frac{\partial^2\left(I_{\rm 4}I_{\rm 5}\right)}{\partial E_{\rm23}^2}=0.
\label{A7}\end{eqnarray}
Eventually, we obtain
\begin{eqnarray}
\gamma_{\rm23} & =\mu_T-\left(3\mu_T+\frac{A}{2}-\alpha_3\right)e \nonumber\\
&=\mu_{\rm T}+\frac{1}{E_{\rm L}}\left(3\mu_{\rm T}+\frac{A}{2}-\alpha_3\right)\sigma_{\rm 11}.
\label{gamma23e}
\end{eqnarray}

We may introduce the non-dimensional coefficients of nonlinearity $\beta_\parallel$ and $\beta_\perp$ used in the main paper, as
\begin{eqnarray}
\eqalign{\beta_\parallel &= 1+\frac{1}{E_{\rm L}}\left(\mu_{\rm L}-\mu_{\rm T}+\frac{A}{4}+\alpha_3+\frac{\alpha_5}{2}\right),\label{Beta23}\\
\beta_{\perp}&=\frac{1}{E_{\rm L}}\left(3\mu_{\rm T}+\frac{A}{2}-\alpha_3\right),
\label{Beta13}
}
\end{eqnarray}
to write the \emph{acousto-elasticity equation of (SH) waves} as follows 
\begin{equation}
\rho_{\rm 0}v_{\rm b}^2=\left(\mu_{\rm L}-\beta_\parallel\sigma_{\rm 11}\right)\cos^2\theta
+\left(\mu_{\rm T}+\beta_\perp\sigma_{\rm 11}\right)\sin^2\theta,\label{resultSH}
\end{equation}
Notice that $\alpha_4$ does not appear at all in that expression.

In the \emph{isotropic limit}, we have  $\mu_{\rm L}=\mu_{\rm T}=\mu$ (the second Lam\'e  coefficient),  $E_{\rm L}=E_{\rm T}=3\mu$, and $\alpha_i=0$, so that (\ref{resultSH}) reduces to 
\begin{equation}
\rho_{\rm 0}v_{\rm b}^2=\mu+\left[-\left(1+\frac{A}{12\mu}\right)\cos^2\theta+\left(1+\frac{A}{6\mu}\right)\sin^2\theta\right]\sigma_{\rm 11},
\label{vbIsotropic}
\end{equation}
in line with the known results of acousto-elasticity \citep{Destrade2010third, abiza2012large}.

In the \emph{absence of pre-stress}, $\sigma_{\rm 11}=0$ and (\ref{resultSH}) reduces to  
\begin{eqnarray}
\rho_{\rm 0}v_{\rm b}^2=\mu_{\rm L}\cos^2\theta + \mu_{\rm T}\sin^2\theta,\label{vbnostress}
\end{eqnarray}
as expected \citep{Chadwick1993}. 
{\it In vivo} experiments \citep{Gennisson2003} indeed demonstrate the existence of slow and fast shear waves in the human biceps, as predicted by this relation.

%%%%%%%%%%%%%%%%%%%%%%

\section{Acousto-elasticity of the (SV) wave}

%%%%%%%%%%%%%%%%%%%%%%%

The (SV) mode has already been studied by \cite{Destrade2010second}. Here we write the results in terms of our choice of moduli.

\cite{Destrade2010second} show that the \emph{acousto-elastic equation of (SV) waves} is
\begin{eqnarray}
\rho_{\rm 0}v_{\rm a}^2 = \alpha \cos^4\theta + 2\beta\cos^2\theta \sin^2\theta + \gamma \sin^4\theta,\label{resultva2}
\end{eqnarray}
where the parameters $\alpha$, $\gamma$ and $\beta$ can be written as follows.
Either in terms of $e$, as
\begin{eqnarray}
\fl\alpha=\mu_{\rm L}+\left(E_{\rm L}+\mu_{\rm L}-\mu_{\rm T}+\frac{A}{4}+\alpha_3+\frac{\alpha_5}{2}\right)e,\nonumber \\ 
\fl\beta= -\mu_{\rm L}+\frac{1}{2}\left(E_{\rm L}+\mu_{\rm T}\right)+
\left(\frac{5}{2}E_{\rm L}-\mu_{\rm L}-5\mu_{\rm T}+\frac{A}{4}+4\alpha_3+3\alpha_4+\frac{5}{2}\alpha_5\right)e,\nonumber \\
\fl\gamma=\mu_{\rm L}+\left(\mu_{\rm L}-\mu_{\rm T}+\frac{A}{4}+\alpha_3+\frac{\alpha_5}{2}\right)e, 
\end{eqnarray}
or terms of $\sigma_{\rm 11}$, as
\begin{eqnarray}
\fl\alpha=\mu_{\rm L}-\left[1+\frac{1}{E_{\rm L}}\left(\mu_{\rm L}-\mu_{\rm T}+\frac{A}{4}+\alpha_3+\frac{\alpha_5}{2} \right)\right]\sigma_{11},\nonumber\\ 
\fl\beta= -\mu_{\rm L}+\frac{1}{2}\left(E_{\rm L}+\mu_{\rm T}\right)-
 \frac{1}{2}\left[5+\frac{1}{E_{\rm L}}\left(-2\mu_{\rm L}-10\mu_{\rm T}+\frac{A}{2}+8\alpha_3+6\alpha_4+5\alpha_5 \right)\right]\sigma_{11},\nonumber\\
\fl\gamma=\mu_{\rm L}-\frac{1}{E_{\rm L}}\left(\mu_{\rm L}-\mu_{\rm T}+\frac{A}{4}+\alpha_3+\frac{\alpha_5}{2} \right)\sigma_{11}. \label{gamma}
\end{eqnarray}

Notice that all the moduli present in the third-order expansion (\ref{W}) of the strain energy $W$ appear in the acousto-elasticity equation for (SV) waves (\ref{resultva2}), although $\alpha_4$ disappears in the special cases of the principal waves at $\theta=0, 90^\circ$.

In the \emph{isotropic limit}, we have  $\mu_{\rm L}=\mu_{\rm T}=\mu$, $E_{\rm L}=E_{\rm T}=3\mu$, $\alpha_i=0$, and the relation reduces to
\begin{eqnarray}
\rho_{\rm 0}v_{\rm a}^2 = \mu + \left( 3\mu \cos^2\theta + \frac{A}{4}\right)e 
=\mu-\left(\cos^2\theta+\frac{A}{12\mu}\right)\sigma_{11},
\label{vaIsotropic}
\end{eqnarray}
which recovers known equations when $\theta=0,90^\circ$ \citep{Gennisson2007, Destrade2010third}.

In the \emph{absence of uniaxial stress},  $\sigma_{\rm 11}=0$, and the relation reduces to 
\begin{eqnarray}
\rho_{\rm 0}v_{\rm a}^2 =\mu_{\rm L}+\left(E_{\rm L}+\mu_{\rm T}-4\mu_{\rm L}\right)\sin^2\theta\cos^2\theta,\label{vanostress}
\end{eqnarray}
in agreement with \cite{Li2016} and \cite{Li2020}.


\begin{thebibliography}{49}
\providecommand{\natexlab}[1]{#1}
\providecommand{\url}[1]{\texttt{#1}}
\expandafter\ifx\csname urlstyle\endcsname\relax
  \providecommand{\doi}[1]{doi: #1}\else
  \providecommand{\doi}{doi: \begingroup \urlstyle{rm}\Url}\fi

\bibitem[Adams and Marano(1995)]{Adams1995}
P.~F. Adams and M.~A. Marano.
\newblock Current estimates from the national health interview survey.
\newblock \emph{Vital and Health Statistics. Series 10, Data from the National
  Health Survey}, \penalty0 (193 Pt 1):\penalty0 1, 1995.

\bibitem[Badley et~al.(1994)Badley, Rasooly, and Webster]{Badley1994}
E.~M. Badley, I.~Rasooly, and G.~K. Webster.
\newblock Relative importance of musculoskeletal disorders as a cause of
  chronic health problems, disability, and health care utilization: findings
  from the 1990 {O}ntario {H}ealth {S}urvey.
\newblock \emph{The Journal of Rheumatology}, 21\penalty0 (3):\penalty0
  505--514, 1994.

\bibitem[Bayat et~al.(2019)Bayat, Adabi, Kumar, Gregory, Webb, Denis, Kim,
  Singh, Mynderse, and Husmann]{Bayat2019}
M.~Bayat, S.~Adabi, V.~Kumar, A.~Gregory, J.~Webb, M.~Denis, B.~Kim, A.~Singh,
  L.~Mynderse, and D.~Husmann.
\newblock Acoustoelasticity analysis of transient waves for non-invasive in
  vivo assessment of urinary bladder.
\newblock \emph{Scientific Reports}, 9\penalty0 (1):\penalty0 1--9, 2019.

\bibitem[Bercoff et~al.(2004)Bercoff, Tanter, Muller, and
  Fink]{bercoff2004role}
J.~Bercoff, M.~Tanter, M.~Muller, and M.~Fink.
\newblock The role of viscosity in the impulse diffraction field of elastic
  waves induced by the acoustic radiation force.
\newblock \emph{IEEE transactions on Ultrasonics, Ferroelectrics, and Frequency
  Control}, 51\penalty0 (11):\penalty0 1523--1536, 2004.

\bibitem[Bernal et~al.(2015)Bernal, Chamming’s, Couade, Bercoff, Tanter, and
  Gennisson]{Bernal2015}
M.~Bernal, F.~Chamming’s, M.~Couade, J.~Bercoff, M.~Tanter, and J.-L.
  Gennisson.
\newblock In vivo quantification of the nonlinear shear modulus in breast
  lesions: {F}easibility study.
\newblock \emph{IEEE Transactions on Ultrasonics, Ferroelectrics, and Frequency
  Control}, 63\penalty0 (1):\penalty0 101--109, 2015.

\bibitem[Bied et~al.(2020)Bied, Jourdain, and
  Gennisson]{bied2020acoustoelasticity}
M.~Bied, L.~Jourdain, and J.-L. Gennisson.
\newblock Acoustoelasticity in transverse isotropic soft tissues:
  quantification of muscles' nonlinear elasticity.
\newblock In \emph{2020 IEEE International Ultrasonics Symposium (IUS)}, pages
  1--4. IEEE, 2020.

\bibitem[Bouillard et~al.(2011)Bouillard, Nordez, and Hug]{Bouillard2011}
K.~Bouillard, A.~Nordez, and F.~Hug.
\newblock Estimation of individual muscle force using elastography.
\newblock \emph{PloS One}, 6\penalty0 (12), 2011.

\bibitem[Bouillard et~al.(2014)Bouillard, Jubeau, Nordez, and
  Hug]{Bouillard2014}
K.~Bouillard, M.~Jubeau, A.~Nordez, and F.~Hug.
\newblock Effect of vastus lateralis fatigue on load sharing between quadriceps
  femoris muscles during isometric knee extensions.
\newblock \emph{Journal of Neurophysiology}, 111\penalty0 (4):\penalty0
  768--776, 2014.

\bibitem[Chadwick(1993)]{Chadwick1993}
P.~Chadwick.
\newblock Wave propagation in incompressible transversely isotropic elastic
  media {I}. {H}omogeneous plane waves.
\newblock \emph{Proceedings of the Royal Irish Academy.}, pages 231--253, 1993.

\bibitem[Deffieux et~al.(2008{\natexlab{a}})Deffieux, Gennisson, Tanter, and
  Fink]{Deffieux2008}
T.~Deffieux, J.-L. Gennisson, M.~Tanter, and M.~Fink.
\newblock Assessment of the mechanical properties of the musculoskeletal system
  using 2-{D} and 3-{D} very high frame rate ultrasound.
\newblock \emph{IEEE Transactions on Ultrasonics, Ferroelectrics, and Frequency
  Control}, 55\penalty0 (10):\penalty0 2177--2190, 2008{\natexlab{a}}.

\bibitem[Deffieux et~al.(2008{\natexlab{b}})Deffieux, Montaldo, Tanter, and
  Fink]{Deffieux2008SWS}
T.~Deffieux, G.~Montaldo, M.~Tanter, and M.~Fink.
\newblock Shear wave spectroscopy for in vivo quantification of human soft
  tissues visco-elasticity.
\newblock \emph{IEEE Transactions on Medical Imaging}, 28\penalty0
  (3):\penalty0 313--322, 2008{\natexlab{b}}.

\bibitem[Destrade et~al.(2010{\natexlab{a}})Destrade, Gilchrist, and
  Ogden]{Destrade2010second}
M.~Destrade, M.~D. Gilchrist, and R.~W. Ogden.
\newblock Third-and fourth-order elasticities of biological soft tissues.
\newblock \emph{The Journal of the Acoustical Society of America}, 127\penalty0
  (4):\penalty0 2103--2106, 2010{\natexlab{a}}.

\bibitem[Destrade et~al.(2010{\natexlab{b}})Destrade, Gilchrist, and
  Saccomandi]{Destrade2010third}
M.~Destrade, M.~D. Gilchrist, and G.~Saccomandi.
\newblock Third-and fourth-order constants of incompressible soft solids and
  the acousto-elastic effect.
\newblock \emph{The Journal of the Acoustical Society of America}, 127\penalty0
  (5):\penalty0 2759--2763, 2010{\natexlab{b}}.

\bibitem[D\'hooge et~al.(2000)D\'hooge, Heimdal, Jamal, Kukulski, Bijnens,
  Rademakers, Hatle, Suetens, and Sutherland]{D'hooge2000}
J.~D\'hooge, A.~Heimdal, F.~Jamal, T.~Kukulski, B.~Bijnens, F.~Rademakers,
  L.~Hatle, P.~Suetens, and G.~R. Sutherland.
\newblock Regional strain and strain rate measurements by cardiac ultrasound:
  {P}rinciples, implementation and limitations.
\newblock \emph{European Journal of Echocardiography}, 1\penalty0 (3):\penalty0
  154--170, 2000.

\bibitem[Downs et~al.(2018)Downs, Lee, Yang, Kim, Wang, and
  Konofagou]{Downs2018}
M.~E. Downs, S.~A. Lee, G.~Yang, S.~Kim, Q.~Wang, and E.~E. Konofagou.
\newblock Non-invasive peripheral nerve stimulation via focused ultrasound in
  vivo.
\newblock \emph{Physics in Medicine \& Biology}, 63\penalty0 (3):\penalty0
  035011, 2018.

\bibitem[Eranki et~al.(2013)Eranki, Cortes, Feren{\v{c}}ek, and
  Sikdar]{Eranki2013}
A.~Eranki, N.~Cortes, Z.~G. Feren{\v{c}}ek, and S.~Sikdar.
\newblock A novel application of musculoskeletal ultrasound imaging.
\newblock \emph{Journal of Visualized Experiments)}, \penalty0 (79):\penalty0
  e50595, 2013.

\bibitem[Ford et~al.(1981)Ford, Huxley, and Simmons]{Ford1981}
L.~Ford, A.~Huxley, and R.~Simmons.
\newblock The relation between stiffness and filament overlap in stimulated
  frog muscle fibres.
\newblock \emph{The Journal of Physiology}, 311\penalty0 (1):\penalty0
  219--249, 1981.

\bibitem[Gennisson et~al.(2003)Gennisson, Catheline, Chaffai, and
  Fink]{Gennisson2003}
J.-L. Gennisson, S.~Catheline, S.~Chaffai, and M.~Fink.
\newblock Transient elastography in anisotropic medium: {A}pplication to the
  measurement of slow and fast shear wave speeds in muscles.
\newblock \emph{The Journal of the Acoustical Society of America}, 114\penalty0
  (1):\penalty0 536--541, 2003.

\bibitem[Gennisson et~al.(2007)Gennisson, R{\'e}nier, Catheline, Barri{\`e}re,
  Bercoff, Tanter, and Fink]{Gennisson2007}
J.-L. Gennisson, M.~R{\'e}nier, S.~Catheline, C.~Barri{\`e}re, J.~Bercoff,
  M.~Tanter, and M.~Fink.
\newblock Acoustoelasticity in soft solids: Assessment of the nonlinear shear
  modulus with the acoustic radiation force.
\newblock \emph{The Journal of the Acoustical Society of America}, 122\penalty0
  (6):\penalty0 3211--3219, 2007.

\bibitem[Gennisson et~al.(2010)Gennisson, Deffieux, Mac{\'e}, Montaldo, Fink,
  and Tanter]{Gennisson2010}
J.-L. Gennisson, T.~Deffieux, E.~Mac{\'e}, G.~Montaldo, M.~Fink, and M.~Tanter.
\newblock Viscoelastic and anisotropic mechanical properties of in vivo muscle
  tissue assessed by supersonic shear imaging.
\newblock \emph{Ultrasound in Medicine \& Biology}, 36\penalty0 (5):\penalty0
  789--801, 2010.

\bibitem[Gijsbertse et~al.(2017)Gijsbertse, Goselink, Lassche, Nillesen,
  Sprengers, Verdonschot, van Alfen, and De~Korte]{Gijsbertse2017}
K.~Gijsbertse, R.~Goselink, S.~Lassche, M.~Nillesen, A.~Sprengers,
  N.~Verdonschot, N.~van Alfen, and C.~De~Korte.
\newblock Ultrasound imaging of muscle contraction of the tibialis anterior in
  patients with facioscapulohumeral dystrophy.
\newblock \emph{Ultrasound in Medicine \& Biology}, 43\penalty0 (11):\penalty0
  2537--2545, 2017.

\bibitem[Hug et~al.(2015)Hug, Tucker, Gennisson, Tanter, and Nordez]{Hug2015}
F.~Hug, K.~Tucker, J.-L. Gennisson, M.~Tanter, and A.~Nordez.
\newblock Elastography for muscle biomechanics: {T}oward the estimation of
  individual muscle force.
\newblock \emph{Exercise and Sport Sciences Reviews}, 43\penalty0 (3):\penalty0
  125--133, 2015.

\bibitem[Jacobson et~al.(1996)Jacobson, Lindgren, and
  Sjukdomarna]{Jacobson1996}
L.~Jacobson, B.~Lindgren, and V.~Sjukdomarna.
\newblock What are the costs of illness?
\newblock \emph{Stockholm: Socialstyrelsen (National Board of Health and
  Welfare)}, 1996.

\bibitem[Jiang et~al.(2015)Jiang, Li, Qian, Liang, Destrade, and
  Cao]{jiang2015measuring}
Y.~Jiang, G.~Li, L.-X. Qian, S.~Liang, M.~Destrade, and Y.~Cao.
\newblock Measuring the linear and nonlinear elastic properties of brain tissue
  with shear waves and inverse analysis.
\newblock \emph{Biomechanics and Modeling in Mechanobiology}, 14\penalty0
  (5):\penalty0 1119--1128, 2015.

\bibitem[Kim et~al.(2018)Kim, Hwang, Kim, Lee, and Jeong]{Kim2018}
K.~Kim, H.-J. Hwang, S.-G. Kim, J.-H. Lee, and W.~K. Jeong.
\newblock Can shoulder muscle activity be evaluated with ultrasound shear wave
  elastography?
\newblock \emph{Clinical Orthopaedics and Related Research}, 476\penalty0
  (6):\penalty0 1276, 2018.

\bibitem[Koo et~al.(2014)Koo, Guo, Cohen, and Parker]{Koo2014}
T.~K. Koo, J.-Y. Guo, J.~H. Cohen, and K.~J. Parker.
\newblock Quantifying the passive stretching response of human tibialis
  anterior muscle using shear wave elastography.
\newblock \emph{Clinical Biomechanics}, 29\penalty0 (1):\penalty0 33--39, 2014.

\bibitem[Latorre-Ossa et~al.(2012)Latorre-Ossa, Gennisson, De~Brosses, and
  Tanter]{Latorre2012}
H.~Latorre-Ossa, J.-L. Gennisson, E.~De~Brosses, and M.~Tanter.
\newblock Quantitative imaging of nonlinear shear modulus by combining static
  elastography and shear wave elastography.
\newblock \emph{IEEE Transactions on Ultrasonics, Ferroelectrics, and Frequency
  Control}, 59\penalty0 (4):\penalty0 833--839, 2012.

\bibitem[Li and Cao(2020)]{Li2020}
G.-Y. Li and Y.~Cao.
\newblock Backward {M}ach cone of shear waves induced by a moving force in soft
  anisotropic materials.
\newblock \emph{Journal of the Mechanics and Physics of Solids}, 138:\penalty0
  103896, 2020.

\bibitem[Li et~al.(2016)Li, Zheng, Liu, Destrade, and Cao]{Li2016}
G.-Y. Li, Y.~Zheng, Y.~Liu, M.~Destrade, and Y.~Cao.
\newblock Elastic {C}herenkov effects in transversely isotropic soft
  materials-i: theoretical analysis, simulations and inverse method.
\newblock \emph{Journal of the Mechanics and Physics of Solids}, 96:\penalty0
  388--410, 2016.

\bibitem[Lopata et~al.(2010)Lopata, van Dijk, Pillen, Nillesen, Maas, Thijssen,
  Stegeman, and de~Korte]{Lopata2010}
R.~G. Lopata, J.~P. van Dijk, S.~Pillen, M.~M. Nillesen, H.~Maas, J.~M.
  Thijssen, D.~F. Stegeman, and C.~L. de~Korte.
\newblock Dynamic imaging of skeletal muscle contraction in three orthogonal
  directions.
\newblock \emph{Journal of Applied Physiology}, 109\penalty0 (3):\penalty0
  906--915, 2010.

\bibitem[Loram et~al.(2006)Loram, Maganaris, and Lakie]{Loram2006}
I.~D. Loram, C.~N. Maganaris, and M.~Lakie.
\newblock Use of ultrasound to make noninvasive in vivo measurement of
  continuous changes in human muscle contractile length.
\newblock \emph{Journal of Applied Physiology}, 100\penalty0 (4):\penalty0
  1311--1323, 2006.

\bibitem[Miyatake et~al.(1995)Miyatake, Yamagishi, Tanaka, Uematsu, Yamazaki,
  Mine, Sano, and Hirama]{Miyatake1995}
K.~Miyatake, M.~Yamagishi, N.~Tanaka, M.~Uematsu, N.~Yamazaki, Y.~Mine,
  A.~Sano, and M.~Hirama.
\newblock New method for evaluating left ventricular wall motion by color-coded
  tissue doppler imaging: in vitro and in vivo studies.
\newblock \emph{Journal of the American College of Cardiology}, 25\penalty0
  (3):\penalty0 717--724, 1995.

\bibitem[Nagueh et~al.(1998)Nagueh, Mikati, Kopelen, Middleton, Qui\~nones, and
  Zoghbi]{Nagueh1998}
S.~F. Nagueh, I.~Mikati, H.~A. Kopelen, K.~J. Middleton, M.~A. Qui\~nones, and
  W.~A. Zoghbi.
\newblock Doppler estimation of left ventricular filling pressure in sinus
  tachycardia: {A} new application of tissue doppler imaging.
\newblock \emph{Circulation}, 98\penalty0 (16):\penalty0 1644--1650, 1998.

\bibitem[Nordez and Hug(2010)]{Nordez2010}
A.~Nordez and F.~Hug.
\newblock Muscle shear elastic modulus measured using supersonic shear imaging
  is highly related to muscle activity level.
\newblock \emph{Journal of Applied Physiology}, 108\penalty0 (5):\penalty0
  1389--1394, 2010.

\bibitem[Nordez et~al.(2008)Nordez, Gennisson, Casari, Catheline, and
  Cornu]{Nordez2008}
A.~Nordez, J.~L. Gennisson, P.~Casari, S.~Catheline, and C.~Cornu.
\newblock Characterization of muscle belly elastic properties during passive
  stretching using transient elastography.
\newblock \emph{Journal of Biomechanics}, 41\penalty0 (10):\penalty0
  2305--2311, 2008.

\bibitem[Ogden and Singh(2011)]{Ogden2011}
R.~W. Ogden and B.~Singh.
\newblock Propagation of waves in an incompressible transversely isotropic
  elastic solid with initial stress: Biot revisited.
\newblock \emph{Journal of Mechanics of Materials and Structures}, 6\penalty0
  (1):\penalty0 453--477, 2011.

\bibitem[Otesteanu et~al.(2019)Otesteanu, Chintada, Rominger, Sanabria, and
  Goksel]{Otesteanu2019}
C.~F. Otesteanu, B.~R. Chintada, M.~B. Rominger, S.~J. Sanabria, and O.~Goksel.
\newblock Spectral quantification of nonlinear elasticity using
  acoustoelasticity and shear-wave dispersion.
\newblock \emph{IEEE Transactions on Ultrasonics, Ferroelectrics, and Frequency
  Control}, 66\penalty0 (12):\penalty0 1845--1855, 2019.

\bibitem[Papazoglou et~al.(2006)Papazoglou, Rump, Braun, and
  Sack]{Papazoglou2006}
S.~Papazoglou, J.~Rump, J.~Braun, and I.~Sack.
\newblock Shear wave group velocity inversion in mr elastography of human
  skeletal muscle.
\newblock \emph{Magnetic Resonance in Medicine: An Official Journal of the
  International Society for Magnetic Resonance in Medicine}, 56\penalty0
  (3):\penalty0 489--497, 2006.

\bibitem[Petit et~al.(1990)Petit, Filippi, Emonet-Denand, Hunt, and
  Laporte]{Petit1990}
J.~Petit, G.~Filippi, F.~Emonet-Denand, C.~Hunt, and Y.~Laporte.
\newblock Changes in muscle stiffness produced by motor units of different
  types in peroneus longus muscle of cat.
\newblock \emph{Journal of Neurophysiology}, 63\penalty0 (1):\penalty0
  190--197, 1990.

\bibitem[Reginster and Khaltaev(2002)]{Reginster2002}
J.-Y. Reginster and N.~Khaltaev.
\newblock Introduction and {WHO} perspective on the global burden of
  musculoskeletal conditions.
\newblock \emph{Rheumatology}, 41\penalty0 (suppl\_1):\penalty0 1--2, 2002.

\bibitem[Rouze et~al.(2013)Rouze, Wang, Palmeri, and Nightingale]{Rouze2013}
N.~C. Rouze, M.~H. Wang, M.~L. Palmeri, and K.~R. Nightingale.
\newblock Finite element modeling of impulsive excitation and shear wave
  propagation in an incompressible, transversely isotropic medium.
\newblock \emph{Journal of Biomechanics}, 46\penalty0 (16):\penalty0
  2761--2768, 2013.

\bibitem[Rouze et~al.(2020)Rouze, Palmeri, and Nightingale]{Rouze2020}
N.~C. Rouze, M.~L. Palmeri, and K.~R. Nightingale.
\newblock Tractable calculation of the {G}reen’s tensor for shear wave
  propagation in an incompressible, transversely isotropic material.
\newblock \emph{Physics in Medicine \& Biology}, 65\penalty0 (1):\penalty0
  015014, 2020.

\bibitem[Sarvazyan et~al.(1998)Sarvazyan, Rudenko, Swanson, Fowlkes, and
  Emelianov]{Sarvazyan1998}
A.~P. Sarvazyan, O.~V. Rudenko, S.~D. Swanson, J.~B. Fowlkes, and S.~Y.
  Emelianov.
\newblock Shear wave elasticity imaging: {A} new ultrasonic technology of
  medical diagnostics.
\newblock \emph{Ultrasound in Medicine \& Biology}, 24\penalty0 (9):\penalty0
  1419--1435, 1998.

\bibitem[Storheim and Zwart(2014)]{Storheim2014}
K.~Storheim and J.-A. Zwart.
\newblock Musculoskeletal disorders and the global burden of disease study,
  2014.

\bibitem[Tran et~al.(2016)Tran, Podwojewski, Beillas, Ott\'enio, Voirin,
  Turquier, and Mitton]{Tran2016}
D.~Tran, F.~Podwojewski, P.~Beillas, M.~Ott\'enio, D.~Voirin, F.~Turquier, and
  D.~Mitton.
\newblock Abdominal wall muscle elasticity and abdomen local stiffness on
  healthy volunteers during various physiological activities.
\newblock \emph{Journal of the Mechanical Behavior of Biomedical Materials},
  60:\penalty0 451--459, 2016.

\bibitem[Vos et~al.(2012)Vos, Flaxman, Naghavi, Lozano, Michaud, Ezzati,
  Shibuya, Salomon, Abdalla, Aboyans, et~al.]{Vos2012}
T.~Vos, A.~D. Flaxman, M.~Naghavi, R.~Lozano, C.~Michaud, M.~Ezzati,
  K.~Shibuya, J.~A. Salomon, S.~Abdalla, V.~Aboyans, et~al.
\newblock Years lived with disability (ylds) for 1160 sequelae of 289 diseases
  and injuries 1990--2010: {A} systematic analysis for the {G}lobal {B}urden of
  {D}isease {S}tudy 2010.
\newblock \emph{The Lancet}, 380\penalty0 (9859):\penalty0 2163--2196, 2012.

\bibitem[{WHO ScientificGroup}(2003)]{ScientificGroup2003}
{WHO ScientificGroup}.
\newblock \emph{The burden of musculoskeletal conditions at the start of the
  new millennium: Report of a WHO Scientific Group}, volume 919.
\newblock World Health Organization, 2003.

\bibitem[Woolf and {\AA}kesson(2001)]{Woolf2001}
A.~D. Woolf and K.~{\AA}kesson.
\newblock Understanding the burden of musculoskeletal conditions, 2001.

\bibitem[Yeung et~al.(1998)Yeung, Levinson, Fu, and Parker]{Yeung1998}
F.~Yeung, S.~F. Levinson, D.~Fu, and K.~J. Parker.
\newblock Feature-adaptive motion tracking of ultrasound image sequences using
  a deformable mesh.
\newblock \emph{IEEE Transactions on Medical Imaging}, 17\penalty0
  (6):\penalty0 945--956, 1998.

\end{thebibliography}

\begin{thebibliography}{12}
\providecommand{\natexlab}[1]{#1}
\providecommand{\url}[1]{\texttt{#1}}
\expandafter\ifx\csname urlstyle\endcsname\relax
  \providecommand{\doi}[1]{doi: #1}\else
  \providecommand{\doi}{doi: \begingroup \urlstyle{rm}\Url}\fi

\bibitem[Abiza et~al.(2012)Abiza, Destrade, and Ogden]{abiza2012large}
Z.~Abiza, M.~Destrade, and R.~W. Ogden.
\newblock Large acoustoelastic effect.
\newblock \emph{Wave Motion}, 49\penalty0 (2):\penalty0 364--374, 2012.

\bibitem[Chadwick(1993)]{Chadwick1993}
P.~Chadwick.
\newblock Wave propagation in incompressible transversely isotropic elastic
  media {I}. {H}omogeneous plane waves.
\newblock \emph{Proceedings of the Royal Irish Academy.}, pages 231--253, 1993.

\bibitem[Destrade et~al.(2010{\natexlab{a}})Destrade, Gilchrist, and
  Ogden]{Destrade2010second}
M.~Destrade, M.~D. Gilchrist, and R.~W. Ogden.
\newblock Third-and fourth-order elasticities of biological soft tissues.
\newblock \emph{The Journal of the Acoustical Society of America}, 127\penalty0
  (4):\penalty0 2103--2106, 2010{\natexlab{a}}.

\bibitem[Destrade et~al.(2010{\natexlab{b}})Destrade, Gilchrist, and
  Saccomandi]{Destrade2010third}
M.~Destrade, M.~D. Gilchrist, and G.~Saccomandi.
\newblock Third-and fourth-order constants of incompressible soft solids and
  the acousto-elastic effect.
\newblock \emph{The Journal of the Acoustical Society of America}, 127\penalty0
  (5):\penalty0 2759--2763, 2010{\natexlab{b}}.

\bibitem[Gennisson et~al.(2003)Gennisson, Catheline, Chaffai, and
  Fink]{Gennisson2003}
J.-L. Gennisson, S.~Catheline, S.~Chaffai, and M.~Fink.
\newblock Transient elastography in anisotropic medium: {A}pplication to the
  measurement of slow and fast shear wave speeds in muscles.
\newblock \emph{The Journal of the Acoustical Society of America}, 114\penalty0
  (1):\penalty0 536--541, 2003.

\bibitem[Gennisson et~al.(2007)Gennisson, R{\'e}nier, Catheline, Barri{\`e}re,
  Bercoff, Tanter, and Fink]{Gennisson2007}
J.-L. Gennisson, M.~R{\'e}nier, S.~Catheline, C.~Barri{\`e}re, J.~Bercoff,
  M.~Tanter, and M.~Fink.
\newblock Acoustoelasticity in soft solids: Assessment of the nonlinear shear
  modulus with the acoustic radiation force.
\newblock \emph{The Journal of the Acoustical Society of America}, 122\penalty0
  (6):\penalty0 3211--3219, 2007.

\bibitem[Li and Cao(2020)]{Li2020}
G.-Y. Li and Y.~Cao.
\newblock Backward {M}ach cone of shear waves induced by a moving force in soft
  anisotropic materials.
\newblock \emph{Journal of the Mechanics and Physics of Solids}, 138:\penalty0
  103896, 2020.

\bibitem[Li et~al.(2016)Li, Zheng, Liu, Destrade, and Cao]{Li2016}
G.-Y. Li, Y.~Zheng, Y.~Liu, M.~Destrade, and Y.~Cao.
\newblock Elastic {C}herenkov effects in transversely isotropic soft
  materials-i: theoretical analysis, simulations and inverse method.
\newblock \emph{Journal of the Mechanics and Physics of Solids}, 96:\penalty0
  388--410, 2016.

\bibitem[Ogden and Singh(2011)]{Ogden2011}
R.~W. Ogden and B.~Singh.
\newblock Propagation of waves in an incompressible transversely isotropic
  elastic solid with initial stress: Biot revisited.
\newblock \emph{Journal of Mechanics of Materials and Structures}, 6\penalty0
  (1):\penalty0 453--477, 2011.

\bibitem[Papazoglou et~al.(2006)Papazoglou, Rump, Braun, and
  Sack]{Papazoglou2006}
S.~Papazoglou, J.~Rump, J.~Braun, and I.~Sack.
\newblock Shear wave group velocity inversion in mr elastography of human
  skeletal muscle.
\newblock \emph{Magnetic Resonance in Medicine: An Official Journal of the
  International Society for Magnetic Resonance in Medicine}, 56\penalty0
  (3):\penalty0 489--497, 2006.

\bibitem[Rouze et~al.(2013)Rouze, Wang, Palmeri, and Nightingale]{Rouze2013}
N.~C. Rouze, M.~H. Wang, M.~L. Palmeri, and K.~R. Nightingale.
\newblock Finite element modeling of impulsive excitation and shear wave
  propagation in an incompressible, transversely isotropic medium.
\newblock \emph{Journal of Biomechanics}, 46\penalty0 (16):\penalty0
  2761--2768, 2013.

\bibitem[Rouze et~al.(2020)Rouze, Palmeri, and Nightingale]{Rouze2020}
N.~C. Rouze, M.~L. Palmeri, and K.~R. Nightingale.
\newblock Tractable calculation of the {G}reen’s tensor for shear wave
  propagation in an incompressible, transversely isotropic material.
\newblock \emph{Physics in Medicine \& Biology}, 65\penalty0 (1):\penalty0
  015014, 2020.

\end{thebibliography}
\end{document}